\documentclass[preprint,5p,times,twocolumn,times]{elsarticle}

\usepackage[final]{epsfig}

\usepackage{xspace}
\usepackage{pstricks}

\usepackage{color}
\usepackage{hyperref}
\usepackage{placeins}
\usepackage{amssymb}
\usepackage{amsmath}

\usepackage{lineno}
\newcommand{\beq}  {\begin{equation}}
\newcommand{\eeq}  {\end{equation}}

\biboptions{sort&compress}

\usepackage{color}

\journal{Physics Letter B}

\begin{document}

\begin{frontmatter}



\title{Beam-helicity asymmetries for single-hadron production in semi-inclusive deep-inelastic scattering from unpolarized hydrogen and deuterium targets}
\author[14,17]{A.~Airapetian}
\author[28]{N.~Akopov}
\author[7]{Z.~Akopov}
\author[8]{E.C.~Aschenauer}
\author[27]{W.~Augustyniak}
\author[20]{S.~Belostotski}
\author[19,26]{H.P.~Blok}
\author[21]{V.~Bryzgalov}
\author[12]{G.P.~Capitani}
\author[23]{E.~Cisbani}
\author[11]{G.~Ciullo}
\author[11]{M.~Contalbrigo}
\author[7]{W.~Deconinck}
\author[12]{E.~De~Sanctis}
\author[10]{M.~Diefenthaler}
\author[12]{P.~Di~Nezza}
\author[14]{M.~D\"uren}
\author[28]{G.~Elbakian}
\author[6]{F.~Ellinghaus}
\author[12]{A.~Fantoni}
\author[24]{L.~Felawka}
\author[21]{G.~Gapienko}
\author[23]{F.~Garibaldi}
\author[7,20,24]{G.~Gavrilov}
\author[28]{V.~Gharibyan}
\author[8]{A.~Hillenbrand}
\author[7]{Y.~Holler}
\author[21]{A.~Ivanilov}
\author[1]{H.E.~Jackson\fnref{*}}
\author[13]{S.~Joosten}
\author[15]{R.~Kaiser}
\author[7,28]{G.~Karyan}
\author[6]{E.~Kinney}
\author[20]{A.~Kisselev}
\author[21]{V.~Korotkov\fnref{*}}
\author[18]{V.~Kozlov}
\author[10,20]{P.~Kravchenko}
\author[2]{L.~Lagamba}
\author[19]{L.~Lapik\'as}
\author[15]{I.~Lehmann}
\author[11]{P.~Lenisa}
\author[17]{W.~Lorenzon}
\author[20]{S.I.~Manaenkov}
\author[27]{B.~Marianski}
\author[28]{H.~Marukyan}
\author[11,28]{A.~Movsisyan}
\author[12]{V.~Muccifora}
\author[10]{A.~Nass}
\author[28]{G.~Nazaryan}
\author[8]{W.-D.~Nowak}
\author[11]{L.L.~Pappalardo}
\author[12]{A.R.~Reolon}
\author[8,16]{C.~Riedl}
\author[10]{K.~Rith}
\author[15]{G.~Rosner}
\author[7]{A.~Rostomyan}
\author[13]{D.~Ryckbosch}
\author[4,5,13]{G.~Schnell}
\author[15]{B.~Seitz}
\author[25]{T.-A.~Shibata}
\author[9]{V.~Shutov}
\author[11]{M.~Statera}
\author[18]{A.~Terkulov}
\author[27]{A.~Trzcinski\fnref{*}}
\author[13]{M.~Tytgat}
\author[13]{Y.~Van~Haarlem}
\author[4,13]{C.~Van~Hulse}
\author[20]{D.~Veretennikov}
\author[2]{I.~Vilardi}
\author[10]{C.~Vogel}
\author[10]{S.~Yaschenko}
\author[7,14]{V.~Zagrebelnyy}
\author[10]{D.~Zeiler}
\author[7]{B.~Zihlmann}
\author[27]{P.~Zupranski}

\fntext[*]{deceased}

\address[1]{Physics Division, Argonne National Laboratory, Argonne, Illinois 60439-4843, USA}
\address[2]{Istituto Nazionale di Fisica Nucleare, Sezione di Bari, 70124 Bari, Italy}
\address[4]{Department of Theoretical Physics, University of the Basque Country UPV/EHU, 48080 Bilbao, Spain}
\address[5]{IKERBASQUE, Basque Foundation for Science, 48013 Bilbao, Spain}
\address[6]{Nuclear Physics Laboratory, University of Colorado, Boulder, Colorado 80309-0390, USA}
\address[7]{DESY, 22603 Hamburg, Germany}
\address[8]{DESY, 15738 Zeuthen, Germany}
\address[9]{Joint Institute for Nuclear Research, 141980 Dubna, Russia}
\address[10]{Physikalisches Institut, Universit\"at Erlangen-N\"urnberg, 91058 Erlangen, Germany}
\address[11]{Istituto Nazionale di Fisica Nucleare, Sezione di Ferrara, and Dipartimento di Fisica e Scienze della Terra, Universit\`a di Ferrara, 44122 Ferrara, Italy}
\address[12]{Istituto Nazionale di Fisica Nucleare, Laboratori Nazionali di Frascati, 00044 Frascati, Italy}
\address[13]{Department of Physics and Astronomy, Ghent University, 9000 Gent, Belgium}
\address[14]{II. Physikalisches Institut, Justus-Liebig Universit\"at Gie{\ss}en, 35392 Gie{\ss}en, Germany}
\address[15]{SUPA, School of Physics and Astronomy, University of Glasgow, Glasgow G12 8QQ, United Kingdom}
\address[16]{Department of Physics, University of Illinois, Urbana, Illinois 61801-3080, USA}
\address[17]{Randall Laboratory of Physics, University of Michigan, Ann Arbor, Michigan 48109-1040, USA }
\address[18]{Lebedev Physical Institute, 117924 Moscow, Russia}
\address[19]{National Institute for Subatomic Physics (Nikhef), 1009 DB Amsterdam, The Netherlands}
\address[20]{Petersburg Nuclear Physics Institute, National Research Center Kurchatov Institute, Gatchina, 188300 Leningrad Region, Russia}
\address[21]{Institute for High Energy Physics, National Research Center Kurchatov Institute, Protvino, 142281 Moscow Region, Russia}
\address[23]{Istituto Nazionale di Fisica Nucleare, Sezione di Roma, Gruppo Collegato Sanit\`a, and Istituto Superiore di Sanit\`a, 00161 Roma, Italy}
\address[24]{TRIUMF, Vancouver, British Columbia V6T 2A3, Canada}
\address[25]{Department of Physics, Tokyo Institute of Technology, Tokyo 152, Japan}
\address[26]{Department of Physics and Astronomy, VU University, 1081 HV Amsterdam, The Netherlands}
\address[27]{National Centre for Nuclear Research, 00-689 Warsaw, Poland}
\address[28]{Yerevan Physics Institute, 375036 Yerevan, Armenia\vspace{-0.35cm}}

\author{\phantom{.}\\[1mm] 
 (The HERMES Collaboration)}
\address{}

\begin{abstract}
A measurement of beam-helicity asymmetries for single-hadron production in deep-inelastic scattering is presented. 
Data from the scattering of 27.6 GeV electrons and positrons off 
gaseous hydrogen and deuterium targets were collected by the HERMES experiment. 
The asymmetries are presented separately as a function 
of the Bjorken scaling variable, the hadron transverse momentum, and the fractional energy for 
charged pions and kaons as well as for protons and anti-protons. These asymmetries 
are also presented as a function of the three aforementioned kinematic variables simultaneously.
\end{abstract}


\end{frontmatter}


\section{Introduction}
\label{sec:intro}

The distribution of partons inside a nucleon, $N$, and their fragmentation into hadrons, $h$, have received so far most theoretical and experimental attention at leading twist, i.e., at twist~2 (see Refs.~\cite{Av16,As16,Ba16,Ga16,Ro16} for review).
Using the ``working definition'' of twist, twist $t$ refers to the power suppression $2-t$ of the hard scale of the process through which the nucleon structure 
and hadron formation is studied~\cite{Ja96}. 
Twist-2 parton distribution functions (PDFs) have a probabilistic interpretation in terms of 
finding a parton inside a nucleon as a function of its fraction of the nucleon longitudinal momentum, $x$, in a frame where the latter tends to infinity.
Similarly, twist-2 fragmentation functions (FFs) describe the probability of a parton to fragment into a 
hadron as a function of the hadron fractional energy with respect to that of the gauge boson exchanged in the process. 
Transverse-momentum-dependent (TMD) PDFs and FFs in\-clu\-de in addition the 
dependence on the transverse momentum of respectively the struck parton inside the nucleon and the formed hadron with respect to the direction of the fragmenting parton.  
For PDFs as well as for FFs, spin-independent and spin-dependent configurations of either or both parton and nucleon can be considered, 
and correlations between spin and transverse momentum can be probed.
At leading twist, there exist eight quark TMD PDFs for the spin-$\frac{1}{2}$ nucleon and eight quark single-hadron TMD FFs. 
The majority of these vanishes when integrating over transverse momentum. 

Higher-twist functions, on the other hand, do not have a probabilistic interpretation. These functions involve correlations between quarks and gluons, and as such allow one to probe the quark-gluon dynamics in the non-perturbative regime of quantum chromodynamics (QCD)~\cite{Ja81,Ell83,Ja91a}. 
Also here, there exist spin-averaged and spin-dependent functions~\cite{Ja91b,Go05,Mu95,Ba06,Bo97}, and most of the functions necessarily involve transverse momentum.

\begin{figure}[t]\centering
\includegraphics[width=0.75\linewidth]{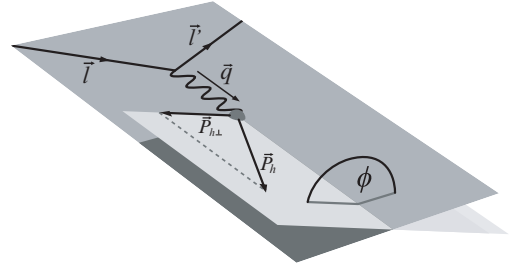}
\caption{\small
Definition of the transverse momentum $\vec{P}_{h\perp}$ and the azimuthal angle $\phi$ of the hadron produced in the semi-inclusive DIS process $e \, N\rightarrow e \, h \, X$.}
\label{fig:kine_plane}
\end{figure}

The decomposition of the single-hadron--production cross section in semi-inclusive deep-inelastic scattering (DIS), $e \, N\rightarrow e\, h \, X$ (where $X$ denotes the generic, unobserved hadronic final state) has been worked out in the one-photon--exchange approximation at 
leading twist and sub-leading twist for small transverse momentum of the produced hadron~\cite{Mu95,Ba06}.  
The cross section is factorized into a hard 
photon-quark scattering process and non-perturbative parts describing the quark distribution and fragmentation~\cite{Col81,Ji04,Col11}.
In this decomposition, the various PDFs and FFs can be isolated pairwise through distinct azimuthal modulations, 
where the azimuthal angles involved are those of the final-state hadron, $\phi$, indicated in Fig.~\ref{fig:kine_plane}, and of the target spin, both 
defined with respect to the lepton-scattering plane about the virtual-photon direction~\cite{Ba04-T}. 
A term independent of the azimuthal angle $\phi$, the beam helicity, and the target spin appears in the cross section at  
twist 2. It provides access to the spin-independent PDF, $f_1$, and the spin-independent 
FF, $D_{1}$, while the leading-twist  $\cos(2\phi)$ modulation, also independent of beam helicity and target spin, provides access to the Boer--Mulders distribution function~\cite{Bo97}, 
$h_1^{\perp}$, in convolution with the Collins FF~\cite{Col93}, $H_{1}^{\perp}$. 
The latter PDF is chiral-odd and describes the distribution
of transversely polarized quarks in an unpolarized nucleon, hereby being sensitive to the correlation between the 
quark transverse momentum and the quark polarization. The Collins FF is also chiral-odd and describes the fragmentation of a transversely polarized quark
into an unpolarized hadron, where the quark polarization generates a distinct angular distribution of the final-state hadron.
Both $h_{1}^{\perp}$ and $H_{1}^{\perp}$ are naive--time-reversal--odd (naive-T-odd), i.e., they change sign when reversing all time coordinates but not the initial
and final states. These functions are found to be non-zero because of final-state interactions~\cite{Col02}.

At sub-leading twist, among others, a $\cos(\phi)$ and a $\sin(\phi)$ modulation appear.
The former is independent of the target and beam polarization state and, in addition to other functions,  
provides access to the Boer-Mulders distribution.
The $\sin(\phi)$ modulation has a term that is sensitive to the beam-helicity and independent of the target-spin state.
The amplitude of this term is proportional to $\sqrt{2\epsilon(1-\epsilon)}\,F^{\sin(\phi)}_{LU}$. 
Here, the subscript $L$ ($U$) stands for longitudinally polarized beam (unpolarized target), and 
\beq
\epsilon=\frac{1-y-\frac{1}{4}\gamma^2 \, y^2}{1-y+\frac{1}{2}y^2+\frac{1}{4}\gamma^2y^2} \nonumber
\eeq
represents the ratio of the fluxes of longitudinally and transversely polarized virtual photons.
The variable $y\equiv P\cdot q/P \cdot l$, with $P$, $q$, and $l$ respectively the four-momenta of the target nucleon, virtual photon, and incoming lepton, represents in the target rest frame the fractional 
virtual-photon energy with respect to that of the incoming lepton, and $\gamma=2Mx/Q$, with $M$ the mass of the target nucleon and $Q^2\equiv-q^2$ the photon virtuality. 
The structure function $F^{\sin(\phi)}_{LU}$ is written at the level of twist 3 in terms of PDFs and FFs as 
\begin{align}
F_{LU}^{\sin(\phi)}(x,Q^2,z,P_{h\perp}) \nonumber \\ 
=\frac{2M}{Q}{\mathcal C}&\left[-\frac{\hat{h}\cdot\vec{k}_{T}}{M_h} \left(xeH^{\perp}_{1}+\frac{M_h}{M}f_1\frac{\tilde{G}^{\perp}}{z}\right) \nonumber \right.\\ 
&\;\,  \left. +\frac{\hat{h}\cdot\vec{p}_{T}}{M}\left( x g^{\perp}D_1+\frac{M_h}{M}h_1^{\perp}\frac{\tilde{E}}{z}\right)\right],
\label{eq:F_LU}
\end{align}
where the mass of the produced hadron is denoted by $M_h$, 
$\hat{h}$ is a unit vector along the direction of the hadron transverse momentum, $\vec{P}_{h\perp}$, 
defined with respect to the virtual-photon direction (as indicated in  Fig.~\ref{fig:kine_plane}), and 
$z$ is in the nucleon rest frame the hadron fractional energy with respect to that of the virtual photon. 
The notation ${\mathcal C} \left[\ldots\right]$ represents a quark-charge--squared weighted sum over (anti-)quarks of convolution 
integrals of the indicated PDFs and FFs~\cite{Ba06}. 
These integrals run over the quark transverse momentum inside the nucleon, $p_T$, and over $k_T$, where $z k_T$ 
represents the hadron transverse momentum acquired during the fragmentation process, relative to the direction of the fragmenting quark.
As indicated in Eq.~(\ref{eq:F_LU}), the structure function depends on the kinematic variables $x$, $Q^2$, $z$, and $P_{h\perp}$. 
The PDFs and FFs also depend on these kinematic variables, but the explicit dependence is omitted. Also for $F^{\sin(\phi)}_{LU}$ this omission is applied in the following. Factorization of structure functions into TMD PDFs and FFs is proven at twist-2~\cite{Col81}. 
For twist-3 structure functions, a factorized decomposition of the form ${\mathcal C} \left[\ldots\right]$ following the twist-2 decomposition is assumed 
on the basis of the parton-model approach, but a proof for this decomposition does not exist. Some issues related to such proof have been discussed in Refs~\cite{Gam06,Eg06,Bo15,Kana15,Chen16}.

From Eq.~(\ref{eq:F_LU}), the twist-3, i.e., $1/Q$ suppression, is apparent. 
The structure function $F_{LU}^{\sin(\phi)}$ is of twist 3 in the low hadron--transverse-momentum region ($P_{h\perp} \ll zQ$). For high values of hadron transverse momentum ($P_{h\perp} \gg M$), the structure function becomes of twist 2 and appears at order $\alpha_S^2$ in the strong coupling constant.\footnote{The hard scale is now given by the transverse momentum and is power-suppressed as $-t$, with $t\ge2$, so that the leading twist still corresponds to twist 2 (see Ref.~\cite{Ba08} for details).} 
In this high--transverse-momentum region, it is described in terms of the 
collinear, i.e., integrated over transverse momentum, 
spin-independent PDF $f_1$ and FF $D_1$, 
and the hadron transverse momentum in the final state finds its origin 
in perturbative QCD radiation~\cite{Ba08}. This is different from the TMD framework, where the hadron  
inherits its transverse momentum from that of the quark inside the nucleon 
and from the fragmentation process. 

Four pairs of PDFs and FFs appear in the 
decomposition of the structure function $F_{LU}^{\sin(\phi)}$ in Eq.~(\ref{eq:F_LU}), where for each pair 
one of the functions is of twist 2, as indicated by the subscript $1$, while the other is of twist 3. 
The functions are also combined such that either the PDF or the FF is naive-T-odd, recognizable here from the superscript $\perp$,
while the other is T-even.
Since the strong interaction conserves chirality, a chiral-odd PDF always appears in pair with a chiral-odd FF, 
e.g., $e$ and $H_{1}^{\perp}$.

The first pair in Eq.~(\ref{eq:F_LU}) is a convolution of the twist-2 Collins FF, mentioned above, and the spin-in\-de\-pen\-dent, twist-3 
PDF $e$~\cite{Ja91b,Ja92}. The genuine twist-3 component of this PDF, following the Wandzura--Wilczek 
decomposition\footnote{For a modern usage of the Wandzura--Wilczek approximation~~\cite{Wa77} in semi-inclusive DIS see Ref.~\cite{Bas18} and references therein.},  
describes the actual  quark-gluon-quark correlations. 
This component can be related to the force exerted on the struck quark by the nucleon remnant at the instant of time that the 
quark absorbs the virtual photon~\cite{Bu08}. 
This interpretation applies to transversely polarized quarks inside an unpolarized nucleon, i.e., for the Boer--Mulders distribution. 
The PDF $e$ is linked via equation-of-motion relations to the spin-independent PDF $f_1$. 
In this relation, the latter appears quark-mass suppressed~\cite{Ef03}. 

The second pair is formed by $f_1$ and the twist-3 chiral-even FF  
$\tilde{G}^{\perp}$~\cite{Ba04}. 
The function $\tilde{G}^{\perp}$ finds its origin in the quark-gluon-quark correlator and vanishes in the Wandzu\-ra--Wilczek approximation.

The following pair contains the chiral-even, twist-2 FF $D_1$, discussed previously, 
and the twist-3 PDF $g^{\perp}$~\cite{Ba04}.
In a similar way as $e$ is linked to $f_1$, $g^{\perp}$ is related to the Boer--Mulders PDF via QCD equations of motion.
A nice property of the $g^{\perp}$ function is that in beam-helicity asymmetries of  
semi-inclusive single-jet production in DIS ($\vec{e} \, N \rightarrow e' \, \text{jet} \, X$, where the symbol $\vec{e}$ indicates a longitudinally polarized lepton beam), it is the only PDF to not vanish, appearing without 
accompanying FF~\cite{Ba06}. 
Semi-inclusive single-jet production provides thus potentially a very clean access to this twist-3 observable.

Finally, the last pair consists of the chiral-odd Boer--Mulders PDF in convolution with the chiral-odd, twist-3 FF 
$\tilde{E}$~\cite{Ja93}. This FF originates from quark-gluon-quark interactions, thus vanishing in the  
Wan\-dzu\-ra--Wilczek approximation. 

The PDFs $e$ and $f_1$ and the FFs $D_1$ and $\tilde{E}$ do not vanish when integrating over
the quark transverse momentum, while the other PDFs and FFs appearing in Eq.~(\ref{eq:F_LU}) vanish.
As can be understood from Eq.~(\ref{eq:F_LU}), the structure function $F_{LU}^{\sin(\phi)}$ goes to zero when integrating over hadron transverse momentum.\footnote{When considering di-hadron production, the analogous structure function $F_{LU}^{\sin(\phi_R)}$ does not vanish when integrating over the individual
hadron transverse momenta, being instead sensitive to the orientation $\phi_R$ of the relative transverse momentum of the two hadrons~\cite{Ba04-R}.}  

The structure function $F_{LU}^{\sin(\phi)}$ can be accessed via the measurement of beam-helicity asymmetries in 
single-hadron production in semi-inclusive DIS off unpolarized hydrogen and deuterium  
targets ($\vec{e} \, N \rightarrow e' \, h \, X$). 
The numerator of these asymmetries contains $F_{LU}^{\sin(\phi)}$, 
while in the denominator a 
convolution of $f_1$ and $D_1$ appears, stemming from the structure function $F_{UU}$, which 
contributes to the spin-independent part of the cross section. 
In addition, there are contributions from the previously mentioned $\cos(\phi)$ and $\cos(2\phi)$ modulations to the denominator.

Very little is known about twist-3 PDFs and FFs.  
There are model-based estimates on $e$~\cite{Me04,Ceb07,Pa18}  
and there are studies on $e$ and its extraction through $eH^{\perp}_{1}$ from experimental data~\cite{Ef02,Ef03,Schw03}, 
extensively treated in Ref.~\cite{Oh03}. 
Another study focusses on predictions of beam-helicity asymmetries involving $h_1^{\perp}\tilde{E}$ only~\cite{Yu03}. 
Evaluations of the combined contributions from $eH^{\perp}_{1}$ and $h_1^{\perp}\tilde{E}$ to the beam-helicity asymmetry 
for positively charged pions, considering scattering off up quarks only, are also available~\cite{Gam04}.
Estimates for $g^{\perp}$ in quark jets are discussed in Ref.~\cite{Af06} and for $g^{\perp}D_1$ in neutral pion production in Ref.~\cite{Ma12}, 
while a combination of 
$g^{\perp}D_1$ and $eH^{\perp}_{1}$ is considered for the estimate of charged-pion and neutral-pion asymmetries~\cite{Ma13}.
Finally, the pair $f_1 \tilde{G}^{\perp}$ is considered and its contribution to the asymmetry 
based on the spectator model is found to be non-negligible~\cite{Ya16}. However, at present there exists no estimate 
for the beam-helicity asymmetry including simultaneously the contributions from all four pairs appearing in Eq.~(\ref{eq:F_LU}). 

Beam-helicity asymmetries were measured by  
CLAS~\cite{CLAS04,CLAS11,CLAS14}, HERMES~\cite{HERMES07}, and COMPASS~\cite{COMPASS14}. 
The CLAS and HERMES experiments provide results for neutral and charged pions, while the COMPASS experiment reports measurements 
for unidentified charged hadrons. 
The HERMES experiment extracted the asymmetries in one dimension, i.e., as a function of one kinematic variable: $z$, $P_{h\perp}$, or  
$x_B$, separately for the low-$z$ ($0.2<z<0.5$) and mid-$z$ ($0.5<z<0.8$) regions. Here, 
$x_B\equiv Q^2/(2P \cdot q)$ can at leading order in $\alpha_s$ be exactly identified with $x$. 
The CLAS experiment performed a two-dimensional extraction in $x_B$ and $P_{h\perp}$. The COMPASS experiment 
assessed three-di\-men\-sional measurements, simultaneously binned in $z$, $P_{h\perp}$, and $x_B$, but only published one-dimensional results because of limited statistical precision.

The present article reports a measurement of the beam-he\-li\-ci\-ty asymmetry by the HERMES experiment for charged pions, 
charged kaons, and (anti-)protons, for data collected on hydrogen and deuterium targets. 
The usage of these two targets is interesting, because it offers different sensitivities to the valence-quark flavors.
This measurement supersedes the former HERMES analysis for charged 
pions~\cite{HERMES07}.\footnote{All the results of Ref.~\cite{HERMES07} 
should hence no longer be used for comparison with other data and in
global fits.}
The asymmetries are extracted from a considerably larger data set. Differently from Ref.~\cite{HERMES07}, 
in the present analysis there is no subtraction of exclusive vector-meson contributions.
The extraction is performed in three dimensions: simultaneously 
in $x_B$, $z$, and $P_{h\perp}$ for pions, kaons, and (anti-)protons as well as in one dimension in each of these three kinematic variables. 

\section{Measurement}
\label{sec:meas}

The analyzed data were collected by the HERMES experiment at DESY between the years 1996 and 2007, from the scattering of 27.6~GeV,  
longitudinally polarized electrons and positrons off pure hydrogen and deuterium gas targets, internal to the HERA lepton storage ring. 
Part of the data were taken with polarized hydrogen or deuterium, but since the data are averaged over the (rapidly reversing) target polarization states, 
they are effectively collected on unpolarized targets. 
The helicity of the lepton beam was reversed roughly every two months, and its polarization value was monitored by two independent Compton polarimeters~\cite{Bar93,Bec00}. 
The average beam-polarization magnitude lies between 34\% and 53\%, depending on the period of data collection.

The HERMES detector consisted of a forward spectrometer divided horizontally into two identical halves above and below the lepton beam~\cite{HERMES98}, 
covering an angular acceptance of $\pm 170$~mrad horizontally and $\pm\left|40-140\right|$~mrad vertically.  
Tracking of electrically charged particles was performed  with the help of a 1.5~Tm dipole magnet and drift chambers 
located upstream and downstream of the magnet, resulting in an average relative momentum resolution of 1.5\%.
Discrimination of leptons and hadrons was performed by a transition-radiation detector, a lead-scintilla\-tor preshower, and a lead-glass electromagnetic calorimeter. 
Typically, a lepton-identi\-fi\-ca\-tion efficiency above 98\% with a misidentification 
of less than 1\% is obtained with this system.
The identification of pions was made possible through the use of a threshold Cherenkov counter in the 
years 1996--1997. This detector was replaced in 1998 by a dual-radiator ring-imaging Cherenkov detector~\cite{Ak00} in order to enable the separation of pions, kaons, and protons.
The trigger system was based on signals from fast hodoscope planes, the preshower, and the calorimeter 
in coincidence with the HERA clock signaling the passage of a beam bunch.

Events are selected based on the presence of a signal formed by adequate responses in the trigger system, 
indicating the possible occurrence of a DIS event. The data also have to meet a series of data-quality requirements, related to proper detector operation, and the beam polarization $P_B$ is restricted to the range $20\% <|P_B|<80\%$.  
From this sample, events for which the kinematic variables reconstructed from the  highest-energy lepton (electron or positron) satisfy the criteria 
$Q^2>1$~GeV$^2$, $0.1<y<0.85$, and $W^2\equiv(P+q)^2>4$~GeV$^2$, where $W$ is the invariant mass of the photon-nucleon system, 
are selected and identified as `DIS events'. 
The criterion on $Q^2$ is a minimal requirement for the selection of the DIS regime. The lower limit on $y$ finds its origin in a degradation of the momentum resolution, while the upper limit is dictated by the trigger threshold, which at the same time
restricts contributions from quantum-electrodynamics (QED) radiation. Finally, a sufficiently high lower limit on $W$ allows one to avoid the kinematic region dominated by electroproduction of baryon resonances.
 
From the sample of DIS events, hadron tracks are selected. The magnitude of their three-momentum $\vec{P}_h$ is restricted to $2(4)~\text{GeV}<|\vec{P}_h|<15$~GeV for pions and kaons (for protons) for the years 1998--2007, and to $4.5~\text{GeV}<|\vec{P}_h|<13.5$~GeV for pions for the years 1996--1997. 
Each such selected hadron track in coincidence with the scattered lepton satisfying the DIS criteria constitutes a `semi-inclusive DIS event'.
The additional requirements $W^2>10~\text{GeV}^2$ and $0.023<x_B<0.4$ are imposed in order to delimit  
a well-defined kinematic phase-space, by excluding regions 
where there are nearly no data present. 
For the same reason, the hadron transverse momentum is limited to $0.05~\text{GeV}<P_{h\perp}<1.8$~GeV.
The hadron fractional energy, $z$, is required to be larger than 0.2 in order to probe current fragmentation, 
and an upper limit of 0.7 is applied in order to reduce contributions from exclusive events, 
as inferred from Monte Carlo simulations. 
In order to study the transition region between semi-inclusive DIS and exclusive processes, 
also results for $z>0.7$ are shown. 

The beam-helicity asymmetry amplitude, $A^{\sin(\phi)}_{LU}$, is extracted from the sample of semi-inclusive DIS events by minimizing the function 
\begin{align}
-\ln \mathbb{L} =  -\sum_i w_i\, \ln \left[1+P_{B,i} \, \sqrt{2\epsilon_i(1-\epsilon_i)} \, A^{\sin(\phi)}_{LU}\,\sin(\phi_i)\right], 
\label{eq:L}
\end{align}
where the sum runs over the semi-inclusive DIS events. 
The weight $w_i$ encodes a correction for erroneously identified DIS events, 
by subtracting events for which the leading lepton is oppositely charged with respect to the beam lepton.
Moreover, the weight $w_i$ assigns, as determined from Monte Carlo simulations, a probability for each identified hadron to be a pion, a kaon, or a proton (see appendix of Ref.~\cite{HERMES10}).
It also effectively equalizes the amount of DIS data collected with positive and negative beam polarization in order to make the normalization constant of the 
probability density function entering the likelihood $\mathbb{L}$ independent of the fit parameters~\cite{An06}.

In addition to the asymmetry amplitude $A^{\sin(\phi)}_{LU}$, the amplitude $\tilde{A}^{\sin(\phi)}_{LU}$ is extracted, by 
removing in the second term of Eq.~(\ref{eq:L}) the prefactor $\sqrt{2\epsilon_i(1-\epsilon_i)}$. This asymmetry effectively includes this prefactor.
The amplitude $A^{\sin(\phi)}_{LU}$ is a measure for the asymmetry with respect to the spin of the virtual photon, 
and hence called {\it virtual-photon asymmetry} in the following, 
while the amplitude $\tilde{A}^{\sin(\phi)}_{LU}$ has the beam-lepton spin as reference, and is referred to as {\it lepton asymmetry}. 
In addition to these amplitudes, a $\sin(2\phi)$ modulation is fit, which is sensitive to the 
two-photon--exchange process~\cite{Schl09}. 

Asymmetries are extracted for each kinematic bin, either in one dimension or in three dimensions, and for each period of stable detector operation. 
For the final result, the asymmetries of each data-taking period are averaged with the inverse of the square of their statistical uncertainty as weight.
This procedure allows one to properly weigh the events according to the beam polarization and to obtain a correct normalization for each period of stable data collection.

Three types of systematic uncertainties are found to contribute to the extracted asymmetry amplitudes.
One originates from the determination of the hadron probability weights.
It is evaluated by using the maximal difference between the central value and the values obtained using 
different Monte Carlo simulations for the evaluation of the hadron weights.  This uncertainty amounts maximally to $5\%$, and on average to $1\%$. Another category of uncertainty stems from QED radiation, finite detector resolution, and limited detector acceptance. 
In order to evaluate these correlated effects, the measured asymmetry is fit with a parametrization depending on 
the kinematic variables $x_B$, $z$, $y$, and $P_{h\perp}$.
This parametrization is implemented in a Monte Carlo simulation that does not include polarization effects. 
The statistical precision of this simulation exceeds that of the experimental data by a factor of ten.
A beam helicity is assigned to each simulated event according to the implemented asymmetry. 
The difference between the implemented asymmetry, evaluated at the average reconstructed kinematics, 
and the one extracted from the fully processed Monte Carlo simulation 
following the same analysis procedure as for experimental data
 is assigned as systematic uncertainty. This uncertainty is the dominant systematic uncertainty.
If the dependence of the asymmetry or the acceptance on the kinematic variables is highly non-linear, 
large differences between the implemented asymmetry and that extracted from the Monte Carlo simulation can arise,
especially for the one-dimensional asymmetries, where one integrates over
a larger portion of phase space. Hence, since the extracted
asymmetries do not exhibit significant non-linear dependences, the
extracted systematic uncertainties due to detector effects are, as
observed, larger for measurements in one dimension than for those in three
dimensions.
In addition, as in general the dependence 
is different for the various kinematic variables, it is natural to obtain from the present procedure 
systematic uncertainties for the various kinematic dependences that differ in size. 
Both systematic uncertainties discussed so far are added in quadrature. 
In addition, a 3\% scale uncertainty from the beam-polarization measurement is assigned.
The influence of additional azimuthal modulations
 [$\sin(2\phi)$, $\cos(\phi)$, and $\cos(2\phi)$, where the first is a beam-helicity--dependent contribution
and the other two are spin-independent contributions to the cross section]
on the extracted $\sin(\phi)$ amplitude was also evaluated, and found to be negligible.
For this study, these modulations were added as additional parameters or existing parameterizations of
the cosine modulations~\cite{HERMES10} were included in the likelihood function.

\section{Results}
\label{sec:result}

\begin{figure*}[!t]\centering
\includegraphics[width=0.92\linewidth]{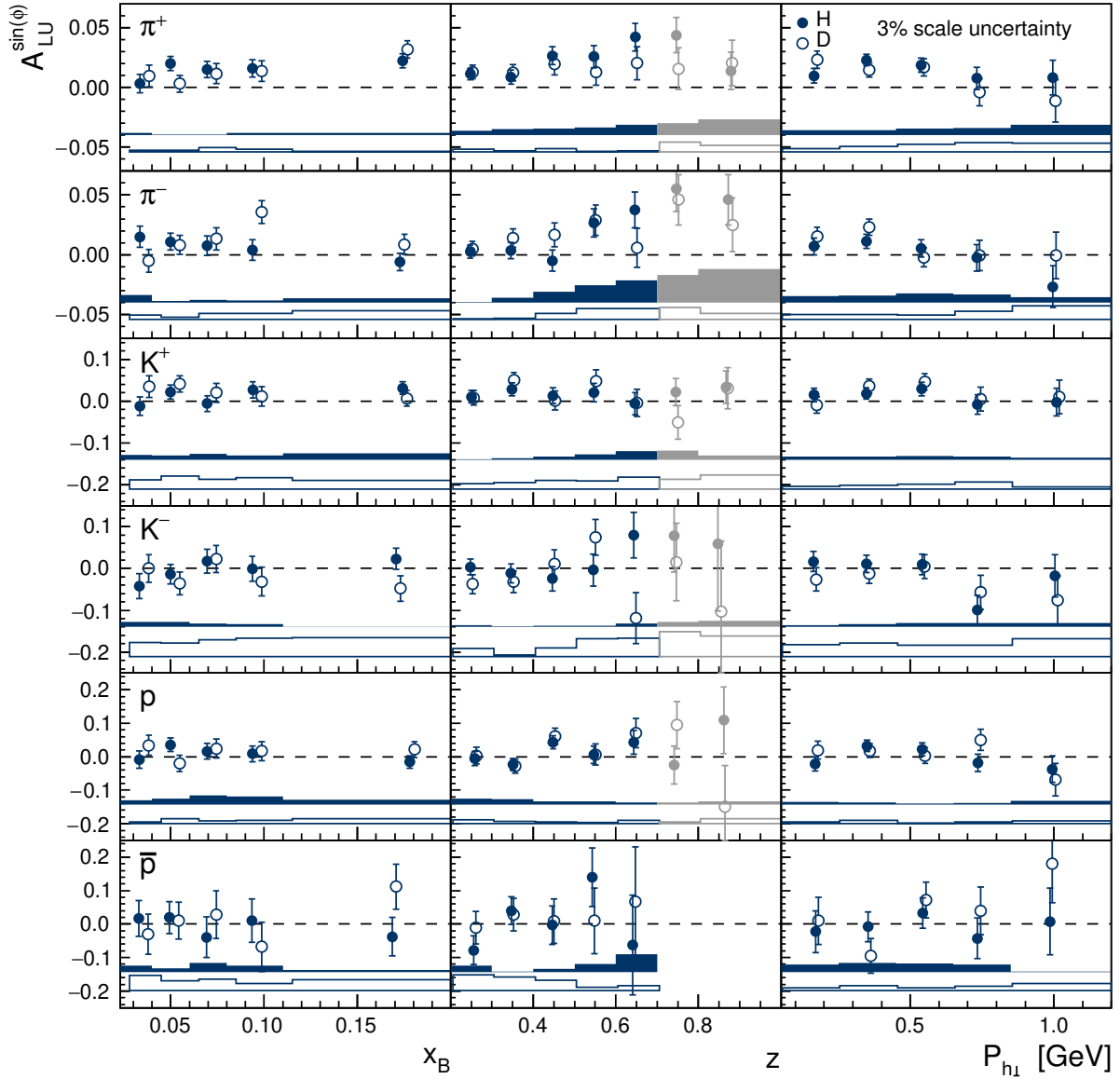}
\caption{\small Virtual-photon asymmetry amplitudes for positively and negatively charged pions 
and kaons, for protons and anti-protons, as a function of $x_{B}$, $z$, and $P_{h\perp}$, for data collected on a hydrogen (closed symbols) and deuterium (open symbols) target. The open symbols are slightly offset horizontally. 
The error bars represent the statistical uncertainties, while the error bands represent systematic uncertainties. 
In addition, there is a systematic uncertainty originating from the measurement of the beam polarization, corresponding to 
a scale factor of 3\%.
The grey data points represent the region for which $z > 0.7$, which is not included in the presentations of the asymmetry amplitudes as a function of $x_{B}$ and $P_{h\perp}$.}
\label{fig:vpa_1d_hadrons}
\end{figure*}

\begin{figure*}\centering
\includegraphics[width=0.7\linewidth]{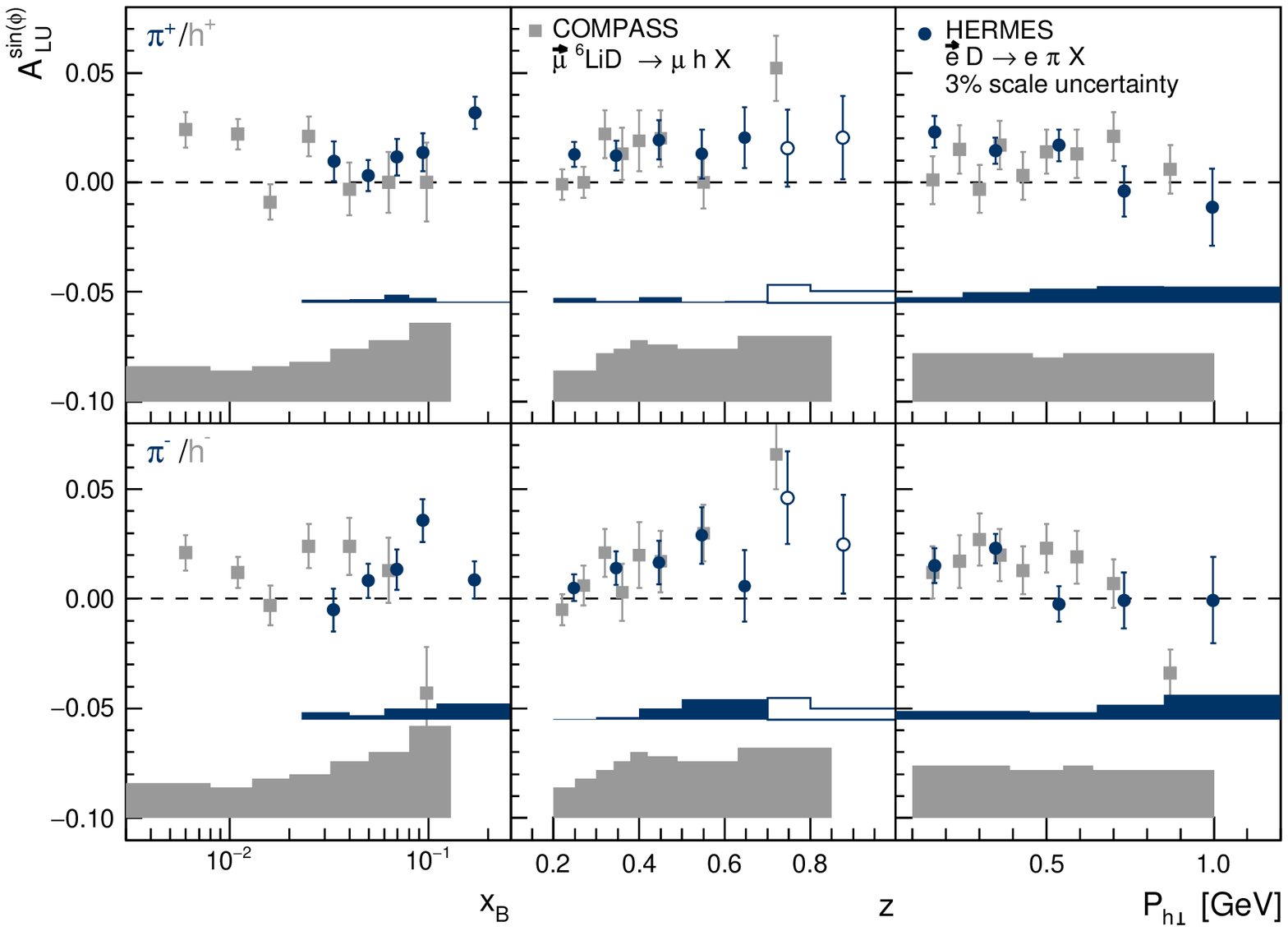}
\caption{\small Virtual-photon asymmetry amplitudes for positively and negatively charged pions, as measured by HERMES on a deuterium target (blue circles), and unidentified hadrons, as measured by COMPASS on a $^6$LiD target (grey squares), as a function of $x_B$, $z$, and $P_{h\perp}$. The open data points from the HERMES measurement represent the region for which $z>0.7$, and are not included in the presentations of the asymmetry amplitudes as a function of $x_B$ and $P_{h\perp}$, while the COMPASS measurement covers the range up to $z = 0.85$ for all projections. The error bars represent the statistical uncertainties, while the error bands represent systematic uncertainties. In addition, there is a systematic uncertainty for the HERMES results 
originating from the measurement of the beam polarization, corresponding to a scale factor of 3\%.}
\label{fig:hermes_compas}
\includegraphics[width=0.7\linewidth]{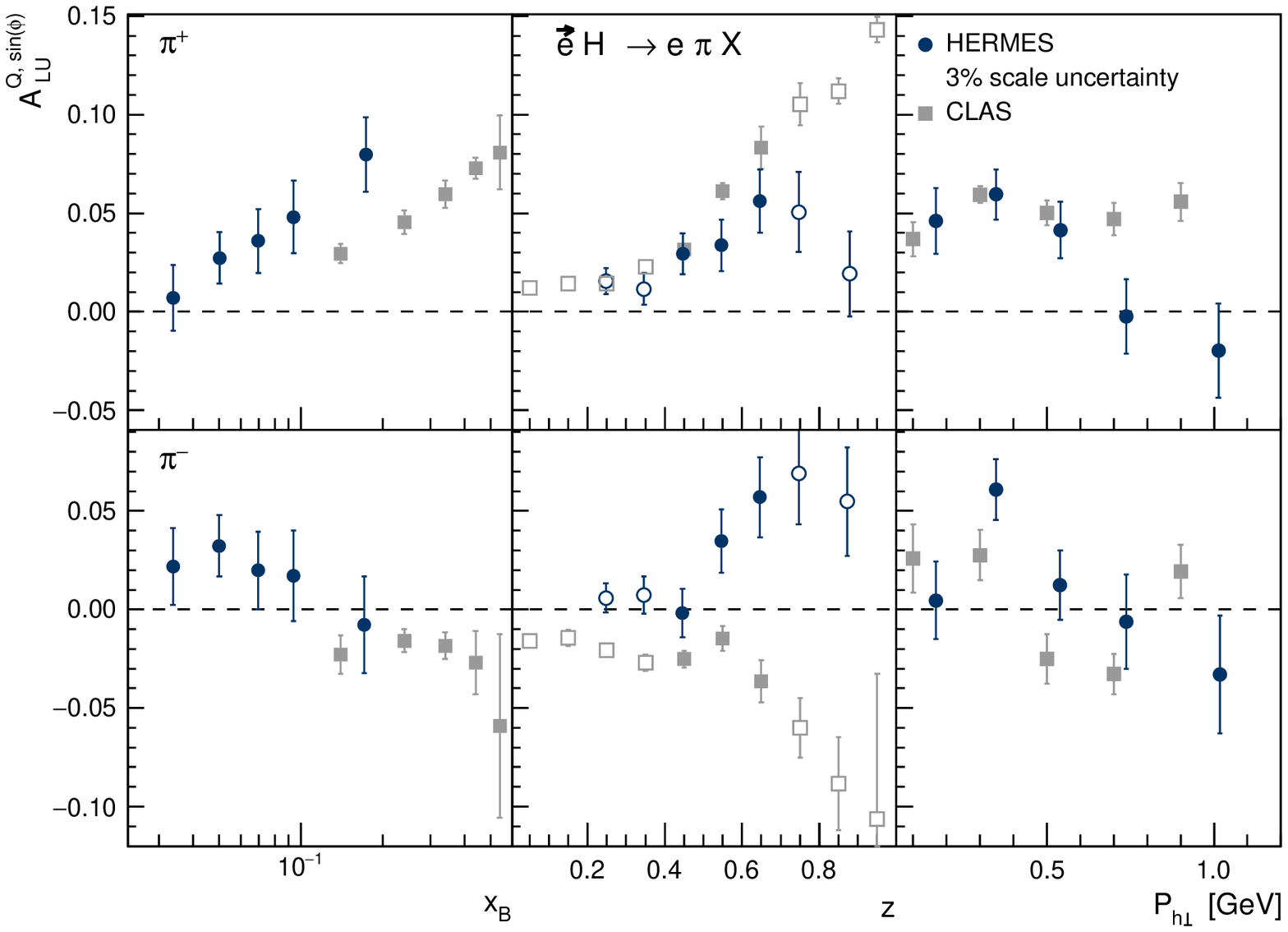}
\caption{\small Virtual-photon asymmetry amplitudes $A_{LU}^{Q,\sin(\phi)}$ for positively and negatively charged pions, as measured by HERMES (blue circles) and CLAS (grey squares) on a hydrogen target, as a function of $x_B$, $z$, and $P_{h\perp}$. 
The data corresponding to the intervals in $z$ indicated by the open symbols are not included in the projections as a function of $x_B$ and $P_{h\perp}$. For both experiments error bars represent the statistical uncertainties only.
There is an additional scale uncertainty of 3\% for the HERMES results originating from the measurement of the beam polarization.} 
\label{fig:hermes_clas}
\end{figure*}

The virtual-photon asymmetry amplitude, $A_{LU}^{\sin(\phi)}$, in one dimension, as a function of $x_B$, $z$, and $P_{h\perp}$, for charge-separated pions and kaons, and for protons and anti-protons is presented in Fig.~\ref{fig:vpa_1d_hadrons} for data collected on a hydrogen target (closed symbols) and on a deuterium target (open symbols). 
The error bars represent the statistical uncertainties, while the error bands are the systematic uncertainties discussed previously. 
As can be seen, the statistical and systematic uncertainties are of the same order of magnitude. 
The data points of the last two bins in $z$, plotted in grey, are not included in the presentations of the asymmetry amplitudes as a function of $x_B$ and $P_{h\perp}$. 
As already stated, the motivation for this lies in the suppression of contributions from exclusive processes.  
Nevertheless, it is interesting to also inspect the asymmetry amplitudes at high $z$, covering the transition region between semi-inclusive DIS and exclusive processes.
The presentation of the asymmetry as a function of $z$ is restricted to $z$ below 0.7 for anti-protons due to limited statistical precision. 

\begin{figure*}[t]\centering
\includegraphics[width=0.9\linewidth]{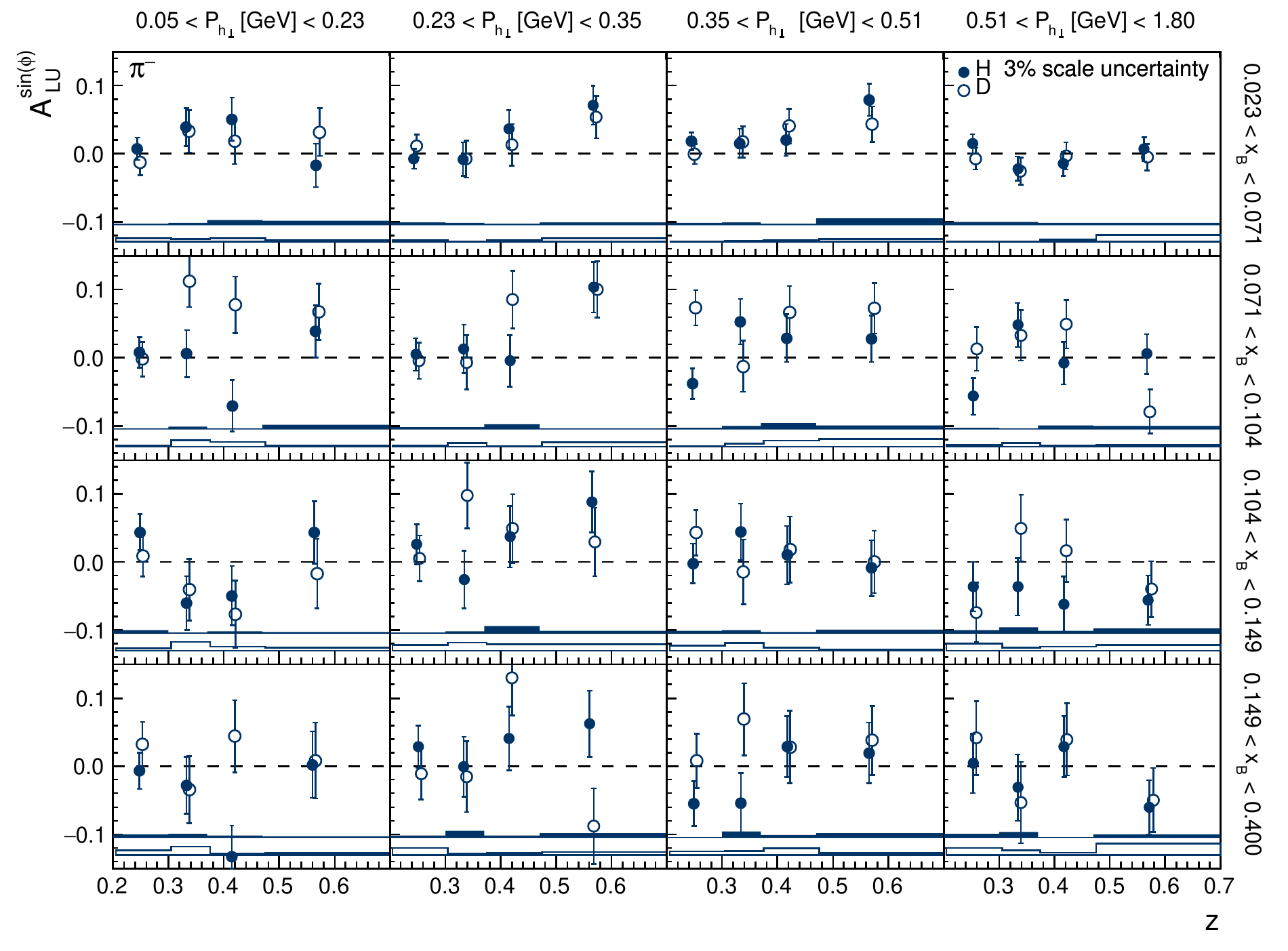}
\caption{\small Virtual-photon asymmetry amplitudes for negatively charged pions as a function of $z$ for slices in $P_{h\perp}$ (columns) and $x_{B}$ (rows), for data collected on a hydrogen (closed symbols) and deuterium (open symbols) target.
The error bars represent the statistical uncertainties, while the error bands represent systematic uncertainties. 
In addition, there is a systematic uncertainty originating from the measurement of the beam polarization, corresponding to 
a scale factor of 3\%.}
\label{fig:vpa_3d_piminus}
\end{figure*}

The virtual-photon asymmetries measured on hydrogen and deuterium targets are in agreement with each other, for all hadron types.
The asymmetry is non-zero for positively and negatively charged pions, with an amplitude rising as a function of $z$. 
For positively charged pions a positive asymmetry amplitude is observed, slightly increasing as a function of $x_B$. An overall positive amplitude
is also seen for negatively charged pions. 
An increase of the amplitude 
as a function of $P_{h\perp}$ for low values of $P_{h\perp}$, followed by a decrease at higher $P_{h\perp}$ values could  
possibly also be distinguished for both pion types. 
For positively charged kaons, a small, positive amplitude is seen, but without any pronounced kinematic dependence, while for negative kaons, protons, and anti-protons the asymmetry amplitudes are compatible with zero.

In Fig.~\ref{fig:hermes_compas}, the virtual-photon asymmetry amplitudes for pions measured on a deuterium target  
are presented together with those for charge-separated hadrons, as measured by the COMPASS experiment on a $^6$LiD target~\cite{COMPASS14}.
Despite the higher average $Q^2$ range of the COMPASS experiment, the two sets of results are compatible with each other.

Another comparison is given in Fig.~\ref{fig:hermes_clas}, 
where results from the analysis discussed here and from the CLAS experiment~\cite{CLAS14} 
for positively and negatively charged pions are shown for data collected on a hydrogen target. 
The CLAS experiment provides data for an asymmetry, $A_{LU}^{Q,\sin(\phi)}$, similar to that defined in Eq.~(\ref{eq:L}), but where the asymmetry 
amplitude in the likelihood function is scaled with $Q$. This is obtained by 
introducing in the second term of Eq.~(\ref{eq:L}) the prefactor $1/Q_i$ and determining its value for each event.
This allows for a comparison between both experimental results free from the 
explicit $1/Q$ factor appearing in Eq.~(\ref{eq:F_LU}). 
The most salient observation 
from this figure is the opposite sign of the asymmetry amplitudes for 
negatively charged pions as a function of $z$ seen by the two experiments.
As can be seen from the projection in $x_B$, the data from the CLAS experiment are located at larger values in $x_B$. 
In the expression for the structure function $F_{LU}^{\sin(\phi)}$ from Eq.~(\ref{eq:F_LU}), the pairs 
$eH^{\perp}_{1}$ and $g^{\perp}D_1$ appear weighted with $x_B$, and thus suppressed at smaller $x_B$ values. 
The Collins FFs for up quarks extracted from data from electron-positron annihilation~\cite{Belle06,Belle08,BABAR13} 
and semi-inclusive DIS~\cite{HERMES04,HERMES10b,HALLA11,COMPASS12} are positive for positively charged pions, and negative for negatively 
charged pions.
Therefore, if $eH^{\perp}_{1}$ forms the dominant contribution to the structure function $F_{LU}^{\sin(\phi)}$ and scattering takes predominantly place 
off an up quark, opposite signs are expected for the asymmetries for positive and negative pions. 
A positive asymmetry is indeed observed for positively charged pions, while a negative asymmetry is seen 
for negatively charged pions for the CLAS measurement, which probes the valence-quark region.
On the other hand, the asymmetries from the present paper, sensitive to lower values of $x_B$, 
are positive for both positively and negatively charged pions. This could hint at the dominance of contributions from 
different pairs of PDFs and FFs appearing in Eq.~(\ref{eq:F_LU}). 

Finally, the virtual-photon asymmetry in three dimensions, as a function of $z$, in bins of $x_B$ and $P_{h\perp}$, is presented in Fig.~\ref{fig:vpa_3d_piminus} for negatively charged pions.
Unlike the one-dimensional results, the uncertainties on the data points are now dominated by the statistical precision.  
The rise of the asymmetry amplitude as a function of $z$ seen in the one-dimensional extraction 
is observed here for certain regions in $x_B$ and $P_{h\perp}$, but within the statistical precision, a delimitation of this behaviour to specific regions in $x_B$ and $P_{h\perp}$ is not observed. 
Also for the other hadrons, no distinctive kinematic dependence is visible for the three-dimensional measurement. 
Similar to the asymmetry measurements in one dimension, no significant differences are observed between the three-dimensional extractions on hydrogen 
and deuterium targets. 
This might be the result of a compensation by the various contributing terms for possible differences
between up and down quarks.

All measured lepton and virtual-photon asymmetries are tabulated in~\ref{tables}. 
Due to the inclusion of the prefactor $\sqrt{2\epsilon_i(1-\epsilon_i)}$, 
the amplitudes of the virtual-photon asymmetry are 
on average larger than those of the 
lepton asymmetry and of the same sign. Nevertheless, with the inclusion
of this prefactor at event level in the fit, the extracted mean value 
of the virtual-photon asymmetry is in certain kinematic bins smaller than
that of the lepton asymmetry and of opposite sign. However, taking into
account the uncertainties, no inconsistencies are observed between the
two asymmetry amplitudes.

The amplitudes of the $\sin(2\phi)$ modulations, sensitive to two-photon--exchange processes, are found to be consistent with zero, within statistical
precision.

\section{Summary and conclusions}

Virtual-photon and lepton asymmetries for char\-ge-separated pions and kaons, and for protons and anti-protons for data 
collected on hydrogen and deuterium targets are presented and discussed. The extraction is performed in one and in three dimensions
in the kinematic variables $x_B$, $z$, and $P_{h\perp}$.

The asymmetries are found to be positive, rising as a function of $z$ for positively and
negatively charged pions, while those for positively charged kaons are found to be slightly positive, but without a specific kinematic
dependence.
The asymmetries for negatively charged kaons, protons, and anti-protons are found to be compatible with zero.
No significant differences are observed between the measurements on hydrogen and deuterium targets.

The virtual-photon asymmetries for pions are found to be in good agreement with the measurement from the 
COMPASS experiment~\cite{COMPASS14}, while a comparison with the results from the CLAS experiment~\cite{CLAS14} suggests a change of sign with increasing
$x_B$ of the asymmetry for negatively charged pions.

The present results constitute the first three-dimensional extraction for charge-separated pions, complementing
the existing one-dimensional and two-dimensional measurements for identified charged pions~\cite{CLAS04,CLAS11,CLAS14, HERMES07} 
and the one-dimensional results for unidentified hadrons~\cite{COMPASS14}.  
For the first time, results for the beam-helicity asymmetry are presented for charged kaons, for protons, and for anti-protons.
The results are presented binned in one dimension and in three dimensions. 
These data can serve therefore as useful input to understand twist-3 PDFs and FFs and quark-gluon-quark  
correlations inside the nucleon and in hadronization, and disentangle the contributions from the various twist-3 PDFs and FFs 
to the beam-helicity asymmetry. 

\section*{Acknowledgments}
We gratefully acknowledge the DESY management for its support, the staff
at DESY and the collaborating institutions for their significant effort.
This work was supported by
the State Committee of Science of the Republic of Armenia, Grant No. 18T-1C180;
the FWO-Flanders and IWT, Belgium;
the Natural Sciences and Engineering Research Council of Canada;
the National Natural Science Foundation of China;
the Alexander von Humboldt Stiftung,
the German Bundesministerium f\"ur Bildung und Forschung (BMBF), and
the Deutsche Forschungsgemeinschaft (DFG);
the Italian Istituto Nazionale di Fisica Nucleare (INFN);
the MEXT, JSPS, and G-COE of Japan;
the Dutch Foundation for Fundamenteel Onderzoek der Materie (FOM);
the Russian Academy of Science and the Russian Federal Agency for
Science and Innovations;
the Basque Foundation for Science (IKERBASQUE), the Basque Government, Grant No. IT956-16, and MINECO (Juan de la Cierva), Spain;
the U.K.~Engineering and Physical Sciences Research Council,
the Science and Technology Facilities Council,
and the Scottish Universities Physics Alliance;
as well as the U.S.~Department of Energy (DOE) and the National Science Foundation (NSF).

\section*{References}

\bibliographystyle{elsarticle-num} 
\bibliography{dc97}


\appendix
\FloatBarrier
\section{Extracted asymmetries and average kinematics}
\label{tables}

See Tables \ref{tab:piplus_1d_hy}-\ref{tab:antip_3d_de}.

\begin{table*}[h]
\footnotesize
\begin{center}
\begin{tabular}{|c|c|c c c c c|r r|}
\hline
\multicolumn{2}{|c|}{} & \multicolumn{5}{|c|}{Average value} & \multicolumn{2}{|c|} {$\pi^+$}\\\hline
 \multicolumn{2}{|c|}{Kinematic bin} & $\langle x_B \rangle$ & $\langle y \rangle$ & $\langle Q^2 \rangle$ & $\langle z \rangle$ & $\langle P_{h\perp} \rangle$ &\multicolumn{1}{|c}{$\tilde{A}^{\sin(\phi)}_{LU}$} & \multicolumn{1}{c|}{$A^{\sin(\phi)}_{LU}$}\\ 
 \multicolumn{2}{|c|}{} &  &  & [GeV$^2$] &  & [GeV] &\multicolumn{1}{|c}{} & \multicolumn{1}{c|}{}\\ \hline
                           & 0.023 -- 0.040 & 0.034 & 0.703 & 1.207 & 0.340 & 0.494 & 0.0019 $\pm$ 0.0053 $\pm$ 0.0009 & 0.0032 $\pm$ 0.0078 $\pm$ 0.0017 \\
                           & 0.040 -- 0.060 & 0.050 & 0.589 & 1.511 & 0.359 & 0.429 & 0.0122 $\pm$ 0.0037 $\pm$ 0.0000 & 0.0200 $\pm$ 0.0060 $\pm$ 0.0007 \\
 $x_{B}$                   & 0.060 -- 0.080 & 0.069 & 0.529 & 1.897 & 0.371 & 0.391 & 0.0093 $\pm$ 0.0040 $\pm$ 0.0007 & 0.0150 $\pm$ 0.0070 $\pm$ 0.0007 \\
                           & 0.080 -- 0.110 & 0.094 & 0.495 & 2.396 & 0.379 & 0.369 & 0.0092 $\pm$ 0.0039 $\pm$ 0.0008 & 0.0159 $\pm$ 0.0071 $\pm$ 0.0017 \\
                           & 0.110 -- 0.400 & 0.174 & 0.460 & 4.110 & 0.383 & 0.351 & 0.0121 $\pm$ 0.0031 $\pm$ 0.0006 & 0.0222 $\pm$ 0.0058 $\pm$ 0.0019 \\
\hline
 \end{tabular}
\end{center}
\caption{One-dimensional lepton asymmetry (column 8) and virtual-photon asymmetry (column 9) for 
positively charged pions as a function of $x_B$, $z$, and $P_{h\perp}$ for data collected on a hydrogen target. 
The bin intervals are given in the first two columns. The average values of the indicated kinematic variables are given in columns 3 to 7. 
For the quoted uncertainties, the first is statistical, while the second is systematic.}
\label{tab:piplus_1d_hy}
\end{table*}

\begin{table*}[h]
\footnotesize
\begin{center}
\begin{tabular}{|c|c|c c c c c|r r|}
\hline
\multicolumn{2}{|c|}{} & \multicolumn{5}{|c|}{Average value} & \multicolumn{2}{|c|} {$\pi^-$}\\\hline
 \multicolumn{2}{|c|}{Kinematic bin} & $\langle x_B \rangle$ & $\langle y \rangle$ & $\langle Q^2 \rangle$ & $\langle z \rangle$ & $\langle P_{h\perp} \rangle$ &\multicolumn{1}{|c}{$\tilde{A}^{\sin(\phi)}_{LU}$} & \multicolumn{1}{c|}{$A^{\sin(\phi)}_{LU}$}\\ 
 \multicolumn{2}{|c|}{} &  &  & [GeV$^2$] &  & [GeV] &\multicolumn{1}{|c}{} & \multicolumn{1}{c|}{}\\ \hline
                          & 0.023 -- 0.040 & 0.033 & 0.705 & 1.206 & 0.335 & 0.491 & 0.0099 $\pm$ 0.0060 $\pm$ 0.0040 & 0.0149 $\pm$ 0.0088 $\pm$ 0.0062 \\
                           & 0.040 -- 0.060 & 0.050 & 0.590 & 1.511 & 0.353 & 0.426 & 0.0061 $\pm$ 0.0043 $\pm$ 0.0008 & 0.0109 $\pm$ 0.0070 $\pm$ 0.0013 \\
 $x_{B}$                   & 0.060 -- 0.080 & 0.069 & 0.530 & 1.900 & 0.364 & 0.388 & 0.0061 $\pm$ 0.0047 $\pm$ 0.0010 & 0.0074 $\pm$ 0.0082 $\pm$ 0.0020 \\
                           & 0.080 -- 0.110 & 0.094 & 0.495 & 2.394 & 0.371 & 0.367 & 0.0033 $\pm$ 0.0047 $\pm$ 0.0005 & 0.0038 $\pm$ 0.0086 $\pm$ 0.0019 \\
                           & 0.110 -- 0.400 & 0.173 & 0.459 & 4.060 & 0.373 & 0.353 & -0.0024 $\pm$ 0.0039 $\pm$ 0.0013 & -0.0062 $\pm$ 0.0072 $\pm$ 0.0034 \\
\hline
 \end{tabular}
\end{center}
\caption{One-dimensional lepton asymmetry (column 8) and virtual-photon asymmetry (column 9) for 
negatively charged pions as a function of $x_B$, $z$, and $P_{h\perp}$ for data collected on a hydrogen target. 
The bin intervals are given in the first two columns. The average values of the indicated kinematic variables are given in columns 3 to 7.
For the quoted uncertainties, the first is statistical, while the second is systematic.}
\label{tab:piminus_1d_hy}
\end{table*}

\begin{table*}[h]
\footnotesize
\begin{center}
\begin{tabular}{|c|c|c c c c c|r r|}
\hline
\multicolumn{2}{|c|}{} & \multicolumn{5}{|c|}{Average value} & \multicolumn{2}{|c|} {$K^+$}\\\hline
 \multicolumn{2}{|c|}{Kinematic bin} & $\langle x_B \rangle$ & $\langle y \rangle$ & $\langle Q^2 \rangle$ & $\langle z \rangle$ & $\langle P_{h\perp} \rangle$ &\multicolumn{1}{|c}{$\tilde{A}^{\sin(\phi)}_{LU}$} & \multicolumn{1}{c|}{$A^{\sin(\phi)}_{LU}$}\\ 
 \multicolumn{2}{|c|}{} &  &  & [GeV$^2$] &  & [GeV] &\multicolumn{1}{|c}{} & \multicolumn{1}{c|}{}\\ \hline
                           & 0.023 -- 0.040 & 0.034 & 0.703 & 1.210 & 0.366 & 0.527 & -0.0086 $\pm$ 0.0148 $\pm$ 0.0079 & -0.0123 $\pm$ 0.0218 $\pm$ 0.0110 \\
                           & 0.040 -- 0.060 & 0.050 & 0.590 & 1.515 & 0.384 & 0.452 & 0.0154 $\pm$ 0.0102 $\pm$ 0.0063 & 0.0214 $\pm$ 0.0166 $\pm$ 0.0101 \\
 $x_{B}$                   & 0.060 -- 0.080 & 0.070 & 0.532 & 1.913 & 0.397 & 0.410 & -0.0032 $\pm$ 0.0108 $\pm$ 0.0090 & -0.0058 $\pm$ 0.0189 $\pm$ 0.0140 \\
                           & 0.080 -- 0.110 & 0.094 & 0.501 & 2.429 & 0.406 & 0.386 & 0.0145 $\pm$ 0.0105 $\pm$ 0.0063 & 0.0269 $\pm$ 0.0190 $\pm$ 0.0106 \\
                           & 0.110 -- 0.400 & 0.174 & 0.472 & 4.217 & 0.413 & 0.364 & 0.0149 $\pm$ 0.0083 $\pm$ 0.0072 & 0.0312 $\pm$ 0.0151 $\pm$ 0.0150 \\
\hline
 \end{tabular}
\end{center}
\caption{One-dimensional lepton asymmetry (column 8) and virtual-photon asymmetry (column 9) for 
positively charged kaons as a function of $x_B$, $z$, and $P_{h\perp}$ for data collected on a hydrogen target. 
The bin intervals are given in the first two columns. The average values of the indicated kinematic variables are given in columns 3 to 7.
For the quoted uncertainties, the first is statistical, while the second is systematic.}
\label{tab:Kplus_1d_hy}
\end{table*}

\begin{table*}[h]
\footnotesize
\begin{center}
\begin{tabular}{|c|c|c c c c c|r r|}
\hline
\multicolumn{2}{|c|}{} & \multicolumn{5}{|c|}{Average value} & \multicolumn{2}{|c|} {$K^-$}\\\hline
 \multicolumn{2}{|c|}{Kinematic bin} & $\langle x_B \rangle$ & $\langle y \rangle$ & $\langle Q^2 \rangle$ & $\langle z \rangle$ & $\langle P_{h\perp} \rangle$ &\multicolumn{1}{|c}{$\tilde{A}^{\sin(\phi)}_{LU}$} & \multicolumn{1}{c|}{$A^{\sin(\phi)}_{LU}$}\\ 
 \multicolumn{2}{|c|}{} &  &  & [GeV$^2$] &  & [GeV] &\multicolumn{1}{|c}{} & \multicolumn{1}{c|}{}\\ \hline
                           & 0.023 -- 0.040 & 0.033 & 0.708 & 1.211 & 0.343 & 0.513 & -0.0301 $\pm$ 0.0199 $\pm$ 0.0096 & -0.0422 $\pm$ 0.0293 $\pm$ 0.0124 \\
                           & 0.040 -- 0.060 & 0.050 & 0.598 & 1.532 & 0.356 & 0.437 & -0.0054 $\pm$ 0.0148 $\pm$ 0.0096 & -0.0135 $\pm$ 0.0238 $\pm$ 0.0129 \\
 $x_{B}$                   & 0.060 -- 0.080 & 0.069 & 0.543 & 1.947 & 0.363 & 0.398 & 0.0078 $\pm$ 0.0165 $\pm$ 0.0057 & 0.0179 $\pm$ 0.0284 $\pm$ 0.0080 \\
                           & 0.080 -- 0.110 & 0.094 & 0.513 & 2.481 & 0.369 & 0.375 & 0.0010 $\pm$ 0.0168 $\pm$ 0.0048 & -0.0004 $\pm$ 0.0296 $\pm$ 0.0066 \\
                           & 0.110 -- 0.400 & 0.171 & 0.479 & 4.182 & 0.369 & 0.363 & 0.0085 $\pm$ 0.0143 $\pm$ 0.0007 & 0.0231 $\pm$ 0.0256 $\pm$ 0.0009 \\
\hline
 \end{tabular}
\end{center}
\caption{One-dimensional lepton asymmetry (column 8) and virtual-photon asymmetry (column 9) for 
negatively charged kaons as a function of $x_B$, $z$, and $P_{h\perp}$ for data collected on a hydrogen target. 
The bin intervals are given in the first two columns. The average values of the indicated kinematic variables are given in columns 3 to 7.
For the quoted uncertainties, the first is statistical, while the second is systematic.}
\label{tab:Kminus_1d_hy}
\end{table*}

\begin{table*}[h]
\footnotesize
\begin{center}
\begin{tabular}{|c|c|c c c c c|r r|}
\hline
\multicolumn{2}{|c|}{} & \multicolumn{5}{|c|}{Average value} & \multicolumn{2}{|c|} {$p$}\\\hline
 \multicolumn{2}{|c|}{Kinematic bin} & $\langle x_B \rangle$ & $\langle y \rangle$ & $\langle Q^2 \rangle$ & $\langle z \rangle$ & $\langle P_{h\perp} \rangle$ &\multicolumn{1}{|c}{$\tilde{A}^{\sin(\phi)}_{LU}$} & \multicolumn{1}{c|}{$A^{\sin(\phi)}_{LU}$}\\ 
 \multicolumn{2}{|c|}{} &  &  & [GeV$^2$] &  & [GeV] &\multicolumn{1}{|c}{} & \multicolumn{1}{c|}{}\\ \hline
                           & 0.023 -- 0.040 & 0.033 & 0.708 & 1.211 & 0.363 & 0.539 & -0.0055 $\pm$ 0.0174 $\pm$ 0.0094 & -0.0083 $\pm$ 0.0256 $\pm$ 0.0130 \\
                           & 0.040 -- 0.060 & 0.050 & 0.603 & 1.546 & 0.404 & 0.483 & 0.0222 $\pm$ 0.0125 $\pm$ 0.0117 & 0.0356 $\pm$ 0.0202 $\pm$ 0.0175 \\
 $x_{B}$                   & 0.060 -- 0.080 & 0.069 & 0.552 & 1.980 & 0.430 & 0.448 & 0.0075 $\pm$ 0.0137 $\pm$ 0.0174 & 0.0164 $\pm$ 0.0236 $\pm$ 0.0279 \\
                           & 0.080 -- 0.110 & 0.094 & 0.527 & 2.553 & 0.445 & 0.424 & 0.0039 $\pm$ 0.0134 $\pm$ 0.0133 & 0.0089 $\pm$ 0.0238 $\pm$ 0.0237 \\
                           & 0.110 -- 0.400 & 0.178 & 0.493 & 4.476 & 0.462 & 0.395 & -0.0067 $\pm$ 0.0103 $\pm$ 0.0079 & -0.0157 $\pm$ 0.0187 $\pm$ 0.0153 \\
\hline
 \end{tabular}
\end{center}
\caption{One-dimensional lepton asymmetry (column 8) and virtual-photon asymmetry (column 9) for 
protons as a function of $x_B$, $z$, and $P_{h\perp}$ for data collected on a hydrogen target. 
The bin intervals are given in the first two columns. The average values of the indicated kinematic variables are given in columns 3 to 7.
For the quoted uncertainties, the first is statistical, while the second is systematic.}
\label{tab:p_1d_hy}
\end{table*}

\begin{table*}[h]
\footnotesize
\begin{center}
\begin{tabular}{|c|c|c c c c c|r r|}
\hline
\multicolumn{2}{|c|}{} & \multicolumn{5}{|c|}{Average value} & \multicolumn{2}{|c|} {$\bar{p}$}\\\hline
 \multicolumn{2}{|c|}{Kinematic bin} & $\langle x_B \rangle$ & $\langle y \rangle$ & $\langle Q^2 \rangle$ & $\langle z \rangle$ & $\langle P_{h\perp} \rangle$ &\multicolumn{1}{|c}{$\tilde{A}^{\sin(\phi)}_{LU}$} & \multicolumn{1}{c|}{$A^{\sin(\phi)}_{LU}$}\\ 
 \multicolumn{2}{|c|}{} &  &  & [GeV$^2$] &  & [GeV] &\multicolumn{1}{|c}{} & \multicolumn{1}{c|}{}\\ \hline
                           & 0.023 -- 0.040 & 0.033 & 0.724 & 1.227 & 0.343 & 0.527 & 0.0098 $\pm$ 0.0362 $\pm$ 0.0130 & 0.0169 $\pm$ 0.0530 $\pm$ 0.0183 \\
                           & 0.040 -- 0.060 & 0.049 & 0.647 & 1.645 & 0.368 & 0.491 & 0.0158 $\pm$ 0.0312 $\pm$ 0.0084 & 0.0195 $\pm$ 0.0481 $\pm$ 0.0109 \\
 $x_{B}$                   & 0.060 -- 0.080 & 0.069 & 0.619 & 2.214 & 0.380 & 0.468 & -0.0340 $\pm$ 0.0384 $\pm$ 0.0183 & -0.0393 $\pm$ 0.0610 $\pm$ 0.0268 \\
                           & 0.080 -- 0.110 & 0.094 & 0.600 & 2.902 & 0.384 & 0.450 & 0.0055 $\pm$ 0.0403 $\pm$ 0.0128 & 0.0103 $\pm$ 0.0645 $\pm$ 0.0195 \\
                           & 0.110 -- 0.400 & 0.169 & 0.562 & 4.833 & 0.393 & 0.418 & -0.0232 $\pm$ 0.0344 $\pm$ 0.0060 & -0.0376 $\pm$ 0.0569 $\pm$ 0.0055 \\
\hline
 \end{tabular}
\end{center}
\caption{One-dimensional lepton asymmetry (column 8) and virtual-photon asymmetry (column 9) for 
anti-protons as a function of $x_B$, $z$, and $P_{h\perp}$ for data collected on a hydrogen target. 
The bin intervals are given in the first two columns. The average values of the indicated kinematic variables are given in columns 3 to 7.
For the quoted uncertainties, the first is statistical, while the second is systematic.}
\label{tab:antip_1d_hy}
\end{table*}

\begin{table*}[h]
\footnotesize
\begin{center}
\begin{tabular}{|c|c|c c c c c|r r|}
\hline
\multicolumn{2}{|c|}{} & \multicolumn{5}{|c|}{Average value} & \multicolumn{2}{|c|} {$\pi^+$}\\\hline
 \multicolumn{2}{|c|}{Kinematic bin} & $\langle x_B \rangle$ & $\langle y \rangle$ & $\langle Q^2 \rangle$ & $\langle z \rangle$ & $\langle P_{h\perp} \rangle$ &\multicolumn{1}{|c}{$\tilde{A}^{\sin(\phi)}_{LU}$} & \multicolumn{1}{c|}{$A^{\sin(\phi)}_{LU}$}\\ 
 \multicolumn{2}{|c|}{} &  &  & [GeV$^2$] &  & [GeV] &\multicolumn{1}{|c}{} & \multicolumn{1}{c|}{}\\ \hline
                           & 0.023 -- 0.040 & 0.033 & 0.704 & 1.207 & 0.341 & 0.490 & 0.0063 $\pm$ 0.0061 $\pm$ 0.0012 & 0.0097 $\pm$ 0.0091 $\pm$ 0.0014 \\
                           & 0.040 -- 0.060 & 0.050 & 0.589 & 1.511 & 0.361 & 0.426 & 0.0017 $\pm$ 0.0044 $\pm$ 0.0002 & 0.0031 $\pm$ 0.0071 $\pm$ 0.0016 \\
 $x_{B}$                   & 0.060 -- 0.080 & 0.069 & 0.528 & 1.895 & 0.373 & 0.388 & 0.0049 $\pm$ 0.0047 $\pm$ 0.0010 & 0.0115 $\pm$ 0.0084 $\pm$ 0.0036 \\
                           & 0.080 -- 0.110 & 0.094 & 0.494 & 2.390 & 0.381 & 0.367 & 0.0062 $\pm$ 0.0047 $\pm$ 0.0005 & 0.0137 $\pm$ 0.0087 $\pm$ 0.0024 \\
                           & 0.110 -- 0.400 & 0.172 & 0.460 & 4.049 & 0.385 & 0.351 & 0.0186 $\pm$ 0.0040 $\pm$ 0.0002 & 0.0317 $\pm$ 0.0073 $\pm$ 0.0005 \\
\hline
 \end{tabular}
\end{center}
\caption{One-dimensional lepton asymmetry (column 8) and virtual-photon asymmetry (column 9) for 
positively charged pions as a function of $x_B$, $z$, and $P_{h\perp}$ for data collected on a deuterium target. 
The bin intervals are given in the first two columns. The average values of the indicated kinematic variables are given in columns 3 to 7.
For the quoted uncertainties, the first is statistical, while the second is systematic.}
\label{tab:piplus_1d_de}
\end{table*}

\begin{table*}[h]
\footnotesize
\begin{center}
\begin{tabular}{|c|c|c c c c c|r r|}
\hline
\multicolumn{2}{|c|}{} & \multicolumn{5}{|c|}{Average value} & \multicolumn{2}{|c|} {$\pi^-$}\\\hline
 \multicolumn{2}{|c|}{Kinematic bin} & $\langle x_B \rangle$ & $\langle y \rangle$ & $\langle Q^2 \rangle$ & $\langle z \rangle$ & $\langle P_{h\perp} \rangle$ &\multicolumn{1}{|c}{$\tilde{A}^{\sin(\phi)}_{LU}$} & \multicolumn{1}{c|}{$A^{\sin(\phi)}_{LU}$}\\ 
 \multicolumn{2}{|c|}{} &  &  & [GeV$^2$] &  & [GeV] &\multicolumn{1}{|c}{} & \multicolumn{1}{c|}{}\\ \hline
                           & 0.023 -- 0.040 & 0.033 & 0.705 & 1.206 & 0.338 & 0.490 & -0.0032 $\pm$ 0.0066 $\pm$ 0.0025 & -0.0051 $\pm$ 0.0097 $\pm$ 0.0033 \\
                           & 0.040 -- 0.060 & 0.050 & 0.590 & 1.510 & 0.356 & 0.425 & 0.0062 $\pm$ 0.0048 $\pm$ 0.0014 & 0.0082 $\pm$ 0.0078 $\pm$ 0.0019 \\
 $x_{B}$                   & 0.060 -- 0.080 & 0.069 & 0.528 & 1.893 & 0.368 & 0.386 & 0.0081 $\pm$ 0.0053 $\pm$ 0.0030 & 0.0133 $\pm$ 0.0093 $\pm$ 0.0049 \\
                           & 0.080 -- 0.110 & 0.094 & 0.493 & 2.384 & 0.376 & 0.366 & 0.0187 $\pm$ 0.0053 $\pm$ 0.0033 & 0.0357 $\pm$ 0.0097 $\pm$ 0.0050 \\
                           & 0.110 -- 0.400 & 0.170 & 0.458 & 3.995 & 0.378 & 0.353 & 0.0037 $\pm$ 0.0046 $\pm$ 0.0042 & 0.0086 $\pm$ 0.0085 $\pm$ 0.0073 \\
\hline
 \end{tabular}
\end{center}
\caption{One-dimensional lepton asymmetry (column 8) and virtual-photon asymmetry (column 9) for 
negatively charged pions as a function of $x_B$, $z$, and $P_{h\perp}$ for data collected on a deuterium target. 
The bin intervals are given in the first two columns. The average values of the indicated kinematic variables are given in columns 3 to 7.
For the quoted uncertainties, the first is statistical, while the second is systematic.}
\label{tab:piminus_1d_de}
\end{table*}

\begin{table*}[h]
\footnotesize
\begin{center}
\begin{tabular}{|c|c|c c c c c|r r|}
\hline
\multicolumn{2}{|c|}{} & \multicolumn{5}{|c|}{Average value} & \multicolumn{2}{|c|} {$K^+$}\\\hline
 \multicolumn{2}{|c|}{Kinematic bin} & $\langle x_B \rangle$ & $\langle y \rangle$ & $\langle Q^2 \rangle$ & $\langle z \rangle$ & $\langle P_{h\perp} \rangle$ &\multicolumn{1}{|c}{$\tilde{A}^{\sin(\phi)}_{LU}$} & \multicolumn{1}{c|}{$A^{\sin(\phi)}_{LU}$}\\ 
 \multicolumn{2}{|c|}{} &  &  & [GeV$^2$] &  & [GeV] &\multicolumn{1}{|c}{} & \multicolumn{1}{c|}{}\\ \hline
                           & 0.023 -- 0.040 & 0.034 & 0.703 & 1.208 & 0.366 & 0.520 & 0.0235 $\pm$ 0.0174 $\pm$ 0.0164 & 0.0349 $\pm$ 0.0256 $\pm$ 0.0216 \\
                           & 0.040 -- 0.060 & 0.050 & 0.589 & 1.514 & 0.384 & 0.445 & 0.0215 $\pm$ 0.0120 $\pm$ 0.0225 & 0.0415 $\pm$ 0.0195 $\pm$ 0.0310 \\
 $x_{B}$                   & 0.060 -- 0.080 & 0.069 & 0.531 & 1.905 & 0.396 & 0.404 & 0.0090 $\pm$ 0.0129 $\pm$ 0.0167 & 0.0198 $\pm$ 0.0226 $\pm$ 0.0225 \\
                           & 0.080 -- 0.110 & 0.094 & 0.500 & 2.423 & 0.405 & 0.383 & 0.0059 $\pm$ 0.0129 $\pm$ 0.0172 & 0.0109 $\pm$ 0.0233 $\pm$ 0.0264 \\
                           & 0.110 -- 0.400 & 0.171 & 0.471 & 4.144 & 0.412 & 0.364 & 0.0005 $\pm$ 0.0105 $\pm$ 0.0129 & 0.0067 $\pm$ 0.0191 $\pm$ 0.0199 \\
\hline
 \end{tabular}
\end{center}
\caption{One-dimensional lepton asymmetry (column 8) and virtual-photon asymmetry (column 9) for 
positively charged kaons as a function of $x_B$, $z$, and $P_{h\perp}$ for data collected on a deuterium target. 
The bin intervals are given in the first two columns. The average values of the indicated kinematic variables are given in columns 3 to 7.
For the quoted uncertainties, the first is statistical, while the second is systematic.}
\label{tab:Kplus_1d_de}
\end{table*}

\begin{table*}[h]
\footnotesize
\begin{center}
\begin{tabular}{|c|c|c c c c c|r r|}
\hline
\multicolumn{2}{|c|}{} & \multicolumn{5}{|c|}{Average value} & \multicolumn{2}{|c|} {$K^-$}\\\hline
 \multicolumn{2}{|c|}{Kinematic bin} & $\langle x_B \rangle$ & $\langle y \rangle$ & $\langle Q^2 \rangle$ & $\langle z \rangle$ & $\langle P_{h\perp} \rangle$ &\multicolumn{1}{|c}{$\tilde{A}^{\sin(\phi)}_{LU}$} & \multicolumn{1}{c|}{$A^{\sin(\phi)}_{LU}$}\\ 
 \multicolumn{2}{|c|}{} &  &  & [GeV$^2$] &  & [GeV] &\multicolumn{1}{|c}{} & \multicolumn{1}{c|}{}\\ \hline
                           & 0.023 -- 0.040 & 0.033 & 0.708 & 1.211 & 0.348 & 0.509 & 0.0014 $\pm$ 0.0224 $\pm$ 0.0233 & 0.0004 $\pm$ 0.0329 $\pm$ 0.0328 \\
                           & 0.040 -- 0.060 & 0.050 & 0.596 & 1.528 & 0.360 & 0.434 & -0.0229 $\pm$ 0.0166 $\pm$ 0.0214 & -0.0356 $\pm$ 0.0268 $\pm$ 0.0323 \\
 $x_{B}$                   & 0.060 -- 0.080 & 0.069 & 0.541 & 1.937 & 0.370 & 0.395 & 0.0144 $\pm$ 0.0188 $\pm$ 0.0242 & 0.0228 $\pm$ 0.0325 $\pm$ 0.0390 \\
                           & 0.080 -- 0.110 & 0.094 & 0.509 & 2.463 & 0.373 & 0.374 & -0.0151 $\pm$ 0.0194 $\pm$ 0.0259 & -0.0315 $\pm$ 0.0344 $\pm$ 0.0439 \\
                           & 0.110 -- 0.400 & 0.168 & 0.476 & 4.092 & 0.373 & 0.359 & -0.0245 $\pm$ 0.0170 $\pm$ 0.0237 & -0.0477 $\pm$ 0.0305 $\pm$ 0.0446 \\
\hline
 \end{tabular}
\end{center}
\caption{One-dimensional lepton asymmetry (column 8) and virtual-photon asymmetry (column 9) for 
negatively charged kaons as a function of $x_B$, $z$, and $P_{h\perp}$ for data collected on a deuterium target. 
The bin intervals are given in the first two columns. The average values of the indicated kinematic variables are given in columns 3 to 7.
For the quoted uncertainties, the first is statistical, while the second is systematic.}
\label{tab:Kminus_1d_de}
\end{table*}

\begin{table*}[h]
\footnotesize
\begin{center}
\begin{tabular}{|c|c|c c c c c|r r|}
\hline
\multicolumn{2}{|c|}{} & \multicolumn{5}{|c|}{Average value} & \multicolumn{2}{|c|} {$p$}\\\hline
 \multicolumn{2}{|c|}{Kinematic bin} & $\langle x_B \rangle$ & $\langle y \rangle$ & $\langle Q^2 \rangle$ & $\langle z \rangle$ & $\langle P_{h\perp} \rangle$ &\multicolumn{1}{|c}{$\tilde{A}^{\sin(\phi)}_{LU}$} & \multicolumn{1}{c|}{$A^{\sin(\phi)}_{LU}$}\\ 
 \multicolumn{2}{|c|}{} &  &  & [GeV$^2$] &  & [GeV] &\multicolumn{1}{|c}{} & \multicolumn{1}{c|}{}\\ \hline
                           & 0.023 -- 0.040 & 0.033 & 0.708 & 1.210 & 0.361 & 0.539 & 0.0228 $\pm$ 0.0202 $\pm$ 0.0039 & 0.0341 $\pm$ 0.0298 $\pm$ 0.0049 \\
                           & 0.040 -- 0.060 & 0.050 & 0.603 & 1.544 & 0.403 & 0.481 & -0.0120 $\pm$ 0.0149 $\pm$ 0.0093 & -0.0193 $\pm$ 0.0241 $\pm$ 0.0137 \\
 $x_{B}$                   & 0.060 -- 0.080 & 0.069 & 0.550 & 1.977 & 0.429 & 0.447 & 0.0154 $\pm$ 0.0163 $\pm$ 0.0037 & 0.0244 $\pm$ 0.0283 $\pm$ 0.0067 \\
                           & 0.080 -- 0.110 & 0.094 & 0.527 & 2.553 & 0.443 & 0.424 & 0.0072 $\pm$ 0.0163 $\pm$ 0.0050 & 0.0167 $\pm$ 0.0289 $\pm$ 0.0095 \\
                           & 0.110 -- 0.400 & 0.176 & 0.491 & 4.404 & 0.461 & 0.395 & 0.0153 $\pm$ 0.0129 $\pm$ 0.0046 & 0.0220 $\pm$ 0.0234 $\pm$ 0.0134 \\
\hline
 \end{tabular}
\end{center}
\caption{One-dimensional lepton asymmetry (column 8) and virtual-photon asymmetry (column 9) for 
protons as a function of $x_B$, $z$, and $P_{h\perp}$ for data collected on a deuterium target. 
The bin intervals are given in the first two columns. The average values of the indicated kinematic variables are given in columns 3 to 7.
For the quoted uncertainties, the first is statistical, while the second is systematic.}
\label{tab:p_1d_de}
\end{table*}

\begin{table*}[h]
\footnotesize
\begin{center}
\begin{tabular}{|c|c|c c c c c|r r|}
\hline
\multicolumn{2}{|c|}{} & \multicolumn{5}{|c|}{Average value} & \multicolumn{2}{|c|} {$\bar{p}$}\\\hline
 \multicolumn{2}{|c|}{Kinematic bin} & $\langle x_B \rangle$ & $\langle y \rangle$ & $\langle Q^2 \rangle$ & $\langle z \rangle$ & $\langle P_{h\perp} \rangle$ &\multicolumn{1}{|c}{$\tilde{A}^{\sin(\phi)}_{LU}$} & \multicolumn{1}{c|}{$A^{\sin(\phi)}_{LU}$}\\ 
 \multicolumn{2}{|c|}{} &  &  & [GeV$^2$] &  & [GeV] &\multicolumn{1}{|c}{} & \multicolumn{1}{c|}{}\\ \hline
                           & 0.023 -- 0.040 & 0.033 & 0.723 & 1.227 & 0.345 & 0.527 & -0.0210 $\pm$ 0.0411 $\pm$ 0.0329 & -0.0298 $\pm$ 0.0602 $\pm$ 0.0460 \\
                           & 0.040 -- 0.060 & 0.049 & 0.644 & 1.641 & 0.373 & 0.483 & 0.0100 $\pm$ 0.0357 $\pm$ 0.0204 & 0.0101 $\pm$ 0.0554 $\pm$ 0.0288 \\
 $x_{B}$                   & 0.060 -- 0.080 & 0.069 & 0.614 & 2.201 & 0.385 & 0.460 & 0.0119 $\pm$ 0.0446 $\pm$ 0.0220 & 0.0285 $\pm$ 0.0710 $\pm$ 0.0341 \\
                           & 0.080 -- 0.110 & 0.094 & 0.598 & 2.897 & 0.388 & 0.445 & -0.0309 $\pm$ 0.0463 $\pm$ 0.0168 & -0.0680 $\pm$ 0.0741 $\pm$ 0.0221 \\
                           & 0.110 -- 0.400 & 0.166 & 0.560 & 4.738 & 0.397 & 0.414 & 0.0563 $\pm$ 0.0406 $\pm$ 0.0177 & 0.1120 $\pm$ 0.0670 $\pm$ 0.0325 \\
\hline
 \end{tabular}
\end{center}
\caption{One-dimensional lepton asymmetry (column 8) and virtual-photon asymmetry (column 9) for 
anti-protons as a function of $x_B$, $z$, and $P_{h\perp}$ for data collected on a deuterium target. 
The bin intervals are given in the first two columns. The average values of the indicated kinematic variables are given in columns 3 to 7.
For the quoted uncertainties, the first is statistical, while the second is systematic.}
\label{tab:antip_1d_de}
\end{table*}

\begin{table*}[h]
\footnotesize
\begin{center}
\begin{tabular}{|c|c|c c c c c|r|}
\hline
\multicolumn{2}{|c|}{} & \multicolumn{5}{|c|}{Average value} & \multicolumn{1}{|c|} {$\pi^+$}\\\hline
 \multicolumn{2}{|c|}{Kinematic bin} & $\langle x_B \rangle$ & $\langle y \rangle$ & $\langle Q^2 \rangle$ & $\langle z \rangle$ & $\langle P_{h\perp} \rangle$ &\multicolumn{1}{c|}{$A^{Q,\sin(\phi)}_{LU}$}\\ 
 \multicolumn{2}{|c|}{} &  &  & [GeV$^2$] &  & [GeV] &\multicolumn{1}{c|}{}\\ \hline
                           & 0.023 -- 0.040 & 0.034 & 0.703 & 1.207 & 0.340 & 0.494 & 0.0072 $\pm$ 0.0167 \\
                           & 0.040 -- 0.060 & 0.050 & 0.589 & 1.511 & 0.359 & 0.429 & 0.0274 $\pm$ 0.0130 \\
 $x_{B}$                   & 0.060 -- 0.080 & 0.070 & 0.529 & 1.897 & 0.371 & 0.391 & 0.0359 $\pm$ 0.0162 \\
                           & 0.080 -- 0.110 & 0.094 & 0.495 & 2.396 & 0.379 & 0.369 & 0.0481 $\pm$ 0.0184 \\
                           & 0.110 -- 0.400 & 0.173 & 0.460 & 4.110 & 0.383 & 0.351 & 0.0798 $\pm$ 0.0188 \\
\hline
 \end{tabular}
\end{center}
\caption{One-dimensional virtual-photon asymmetry $A^{Q,\sin(\phi)}_{LU}$ (column 8) for 
positively charged pions as a function of $x_B$, $z$, and $P_{h\perp}$ for data collected on a hydrogen target. 
The bin intervals are given in the first two columns. The average values of the indicated kinematic variables are given in columns 3 to 7. 
Only the statistical uncertainty is quoted.}
\label{tab:piplus_1d_hy_Q}
\end{table*}

\begin{table*}[h]
\footnotesize
\begin{center}
\begin{tabular}{|c|c|c c c c c|r|}
\hline
\multicolumn{2}{|c|}{} & \multicolumn{5}{|c|}{Average value} & \multicolumn{1}{|c|} {$\pi^-$}\\\hline
 \multicolumn{2}{|c|}{Kinematic bin} & $\langle x_B \rangle$ & $\langle y \rangle$ & $\langle Q^2 \rangle$ & $\langle z \rangle$ & $\langle P_{h\perp} \rangle$ &\multicolumn{1}{c|}{$A^{Q,\sin(\phi)}_{LU}$}\\ 
 \multicolumn{2}{|c|}{} &  &  & [GeV$^2$] &  & [GeV] &\multicolumn{1}{c|}{}\\ \hline
                           & 0.023 -- 0.040 & 0.034 & 0.705 & 1.206 & 0.335 & 0.491 & 0.0217 $\pm$ 0.0195 \\
                           & 0.040 -- 0.060 & 0.050 & 0.590 & 1.511 & 0.353 & 0.426 & 0.0323 $\pm$ 0.0155 \\
 $x_{B}$                   & 0.060 -- 0.080 & 0.070 & 0.530 & 1.900 & 0.364 & 0.388 & 0.0198 $\pm$ 0.0198 \\
                           & 0.080 -- 0.110 & 0.094 & 0.495 & 2.394 & 0.371 & 0.367 & 0.0172 $\pm$ 0.0230 \\
                           & 0.110 -- 0.400 & 0.171 & 0.459 & 4.060 & 0.373 & 0.353 & -0.0078 $\pm$ 0.0245 \\
\hline
 \end{tabular}
\end{center}
\caption{One-dimensional virtual-photon asymmetry $A^{Q,\sin(\phi)}_{LU}$ (column 8) for negatively 
charged pions as a function of $x_B$, $z$, and $P_{h\perp}$ for data collected on a hydrogen target. 
The bin intervals are given in the first two columns. The average values of the indicated kinematic variables are given in columns 3 to 7. 
Only the statistical uncertainty is quoted.}
\label{tab:piminus_1d_hy_Q}
\end{table*}

\FloatBarrier


\begin{table*}[h]
\footnotesize
\centering
\begin{tabular}{|c c c |c c c c c|r r|}
\hline
\multicolumn{3}{|c|}{Kinematic bin} & \multicolumn{5}{|c|}{Average value} & \multicolumn{2}{|c|} {$\pi^+$}\\\hline
 $x_{B}$ & $z$ & $P_{h\perp}$ & $\langle x_B \rangle$ & $\langle y \rangle$ & $\langle Q^2 \rangle$ & $\langle z \rangle$ & $\langle P_{h\perp} \rangle$ & \multicolumn{1}{c}{$\tilde{A}^{\sin(\phi)}_{LU}$} & \multicolumn{1}{c|}{$A^{\sin(\phi)}_{LU}$}\\
 & & $\left[\text{GeV}\right]$ &  &  & $\left[\text{GeV}^2\right]$  &  & $\left[\text{GeV}\right]$  &  & \\ \hline
0.023 -- 0.071 & 0.20 -- 0.30 & 0.05 -- 0.23 & 0.052 & 0.560 & 1.460 & 0.244 & 0.158 & 0.0042 $\pm$ 0.0088 $\pm$ 0.0008 & 0.0035 $\pm$ 0.0149 $\pm$ 0.0011 \\
               &              & 0.23 -- 0.35 & 0.049 & 0.614 & 1.505 & 0.244 & 0.292 & 0.0101 $\pm$ 0.0083 $\pm$ 0.0025 & 0.0128 $\pm$ 0.0132 $\pm$ 0.0041 \\
               &              & 0.35 -- 0.51 & 0.047 & 0.645 & 1.530 & 0.245 & 0.426 & 0.0084 $\pm$ 0.0073 $\pm$ 0.0008 & 0.0129 $\pm$ 0.0113 $\pm$ 0.0019 \\
               &              & 0.51 -- 1.80 & 0.045 & 0.686 & 1.572 & 0.252 & 0.633 & 0.0046 $\pm$ 0.0085 $\pm$ 0.0005 & 0.0064 $\pm$ 0.0127 $\pm$ 0.0002 \\
               & 0.30 -- 0.37 & 0.05 -- 0.23 & 0.053 & 0.530 & 1.430 & 0.332 & 0.157 & 0.0032 $\pm$ 0.0141 $\pm$ 0.0008 & 0.0079 $\pm$ 0.0244 $\pm$ 0.0013 \\
\hline
 \end{tabular}
\caption{Three-dimensional lepton asymmetries (column 9) and virtual-photon asymmetries (column 10) for 
positively charged pions as a function of $x_B$, $z$, and $P_{h\perp}$ for data collected on a hydrogen target.
The bin intervals are given in the first three columns. The average values of the indicated kinematic variables are given in columns 4 to 8.
For the quoted uncertainties, the first is statistical, while the second is systematic.}
\label{tab:pi+_3d_hy}
\end{table*}

\begin{table*}[h]
\footnotesize
\centering
\begin{tabular}{|c c c |c c c c c|r r|}
\hline
\multicolumn{3}{|c|}{Kinematic bin} & \multicolumn{5}{|c|}{Average value} & \multicolumn{2}{|c|} {$\pi^-$}\\\hline
 $x_{B}$ & $z$ & $P_{h\perp}$ & $\langle x_B \rangle$ & $\langle y \rangle$ & $\langle Q^2 \rangle$ & $\langle z \rangle$ & $\langle P_{h\perp} \rangle$ & \multicolumn{1}{c}{$\tilde{A}^{\sin(\phi)}_{LU}$} & \multicolumn{1}{c|}{$A^{\sin(\phi)}_{LU}$}\\
 & & $\left[\text{GeV}\right]$ &  &  & $\left[\text{GeV}^2\right]$  &  & $\left[\text{GeV}\right]$  &  & \\ \hline
0.023 -- 0.071 & 0.20 -- 0.30 & 0.05 -- 0.23 & 0.051 & 0.562 & 1.456 & 0.244 & 0.158 & 0.0065 $\pm$ 0.0099 $\pm$ 0.0015 & 0.0070 $\pm$ 0.0167 $\pm$ 0.0020 \\
               &              & 0.23 -- 0.35 & 0.048 & 0.618 & 1.498 & 0.243 & 0.292 & -0.0060 $\pm$ 0.0094 $\pm$ 0.0019 & -0.0079 $\pm$ 0.0149 $\pm$ 0.0037 \\
               &              & 0.35 -- 0.51 & 0.047 & 0.646 & 1.522 & 0.244 & 0.426 & 0.0115 $\pm$ 0.0083 $\pm$ 0.0017 & 0.0185 $\pm$ 0.0128 $\pm$ 0.0027 \\
               &              & 0.51 -- 1.80 & 0.045 & 0.688 & 1.572 & 0.252 & 0.636 & 0.0093 $\pm$ 0.0095 $\pm$ 0.0031 & 0.0144 $\pm$ 0.0141 $\pm$ 0.0046 \\
               & 0.30 -- 0.37 & 0.05 -- 0.23 & 0.053 & 0.531 & 1.425 & 0.332 & 0.157 & 0.0242 $\pm$ 0.0161 $\pm$ 0.0026 & 0.0387 $\pm$ 0.0278 $\pm$ 0.0035 \\
\hline
 \end{tabular}
\caption{Three-dimensional lepton asymmetries (column 9) and virtual-photon asymmetries (column 10) for 
negatively charged pions as a function of $x_B$, $z$, and $P_{h\perp}$ for data collected on a hydrogen target.
The bin intervals are given in the first three columns. The average values of the indicated kinematic variables are given in columns 4 to 8.
For the quoted uncertainties, the first is statistical, while the second is systematic.}
\label{tab:pi-_3d_hy}
\end{table*}

\begin{table*}[h]
\footnotesize
\centering
\begin{tabular}{|c c c |c c c c c|r r|}
\hline
\multicolumn{3}{|c|}{Kinematic bin} & \multicolumn{5}{|c|}{Average value} & \multicolumn{2}{|c|} {$K^+$}\\\hline
 $x_{B}$ & $z$ & $P_{h\perp}$ & $\langle x_B \rangle$ & $\langle y \rangle$ & $\langle Q^2 \rangle$ & $\langle z \rangle$ & $\langle P_{h\perp} \rangle$ & \multicolumn{1}{c}{$\tilde{A}^{\sin(\phi)}_{LU}$} & \multicolumn{1}{c|}{$A^{\sin(\phi)}_{LU}$}\\
 & & $\left[\text{GeV}\right]$ &  &  & $\left[\text{GeV}^2\right]$  &  & $\left[\text{GeV}\right]$  &  & \\ \hline
0.023 -- 0.071 & 0.20 -- 0.30 & 0.05 -- 0.23 & 0.052 & 0.565 & 1.473 & 0.247 & 0.156 & 0.0088 $\pm$ 0.0265 $\pm$ 0.0020 & 0.0147 $\pm$ 0.0435 $\pm$ 0.0035 \\
               &              & 0.23 -- 0.35 & 0.049 & 0.616 & 1.517 & 0.247 & 0.291 & 0.0030 $\pm$ 0.0260 $\pm$ 0.0027 & 0.0017 $\pm$ 0.0409 $\pm$ 0.0041 \\
               &              & 0.35 -- 0.51 & 0.047 & 0.645 & 1.538 & 0.248 & 0.426 & -0.0215 $\pm$ 0.0236 $\pm$ 0.0006 & -0.0228 $\pm$ 0.0360 $\pm$ 0.0008 \\
               &              & 0.51 -- 1.80 & 0.045 & 0.688 & 1.571 & 0.254 & 0.647 & 0.0374 $\pm$ 0.0257 $\pm$ 0.0046 & 0.0573 $\pm$ 0.0382 $\pm$ 0.0063 \\
               & 0.30 -- 0.37 & 0.05 -- 0.23 & 0.054 & 0.533 & 1.442 & 0.333 & 0.156 & 0.0014 $\pm$ 0.0363 $\pm$ 0.0042 & -0.0016 $\pm$ 0.0626 $\pm$ 0.0065 \\
\hline
 \end{tabular}
\caption{Three-dimensional lepton asymmetries (column 9) and virtual-photon asymmetries (column 10) for 
positively charged kaons as a function of $x_B$, $z$, and $P_{h\perp}$ for data collected on a hydrogen target.
The bin intervals are given in the first three columns. The average values of the indicated kinematic variables are given in columns 4 to 8.
For the quoted uncertainties, the first is statistical, while the second is systematic.}
\label{tab:K+_3d_hy}
\end{table*}

\begin{table*}[h]
\footnotesize
\centering
\begin{tabular}{|c c c |c c c c c|r r|}
\hline
\multicolumn{3}{|c|}{Kinematic bin} & \multicolumn{5}{|c|}{Average value} & \multicolumn{2}{|c|} {$K^-$}\\\hline
 $x_{B}$ & $z$ & $P_{h\perp}$ & $\langle x_B \rangle$ & $\langle y \rangle$ & $\langle Q^2 \rangle$ & $\langle z \rangle$ & $\langle P_{h\perp} \rangle$ & \multicolumn{1}{c}{$\tilde{A}^{\sin(\phi)}_{LU}$} & \multicolumn{1}{c|}{$A^{\sin(\phi)}_{LU}$}\\
 & & $\left[\text{GeV}\right]$ &  &  & $\left[\text{GeV}^2\right]$  &  & $\left[\text{GeV}\right]$  &  & \\ \hline
0.023 -- 0.071 & 0.20 -- 0.30 & 0.05 -- 0.23 & 0.051 & 0.569 & 1.469 & 0.245 & 0.157 & 0.0025 $\pm$ 0.0356 $\pm$ 0.0029 & 0.0129 $\pm$ 0.0585 $\pm$ 0.0050 \\
               &              & 0.23 -- 0.35 & 0.048 & 0.624 & 1.517 & 0.244 & 0.292 & -0.0245 $\pm$ 0.0343 $\pm$ 0.0057 & -0.0420 $\pm$ 0.0537 $\pm$ 0.0082 \\
               &              & 0.35 -- 0.51 & 0.047 & 0.651 & 1.530 & 0.246 & 0.426 & -0.0224 $\pm$ 0.0306 $\pm$ 0.0064 & -0.0331 $\pm$ 0.0466 $\pm$ 0.0069 \\
               &              & 0.51 -- 1.80 & 0.045 & 0.693 & 1.579 & 0.252 & 0.648 & -0.0019 $\pm$ 0.0315 $\pm$ 0.0073 & -0.0070 $\pm$ 0.0468 $\pm$ 0.0085 \\
               & 0.30 -- 0.37 & 0.05 -- 0.23 & 0.053 & 0.536 & 1.437 & 0.333 & 0.157 & -0.0848 $\pm$ 0.0527 $\pm$ 0.0025 & -0.1256 $\pm$ 0.0905 $\pm$ 0.0058 \\
\hline
 \end{tabular}
\caption{Three-dimensional lepton asymmetries (column 9) and virtual-photon asymmetries (column 10) for 
negatively charged kaons as a function of $x_B$, $z$, and $P_{h\perp}$ for data collected on a hydrogen target.
The bin intervals are given in the first three columns. The average values of the indicated kinematic variables are given in columns 4 to 8.
For the quoted uncertainties, the first is statistical, while the second is systematic.}
\label{tab:K-_3d_hy}
\end{table*}

\begin{table*}[h]
\footnotesize
\centering
\begin{tabular}{|c c c |c c c c c|r r|}
\hline
\multicolumn{3}{|c|}{Kinematic bin} & \multicolumn{5}{|c|}{Average value} & \multicolumn{2}{|c|} {$p$}\\\hline
 $x_{B}$ & $z$ & $P_{h\perp}$ & $\langle x_B \rangle$ & $\langle y \rangle$ & $\langle Q^2 \rangle$ & $\langle z \rangle$ & $\langle P_{h\perp} \rangle$ & \multicolumn{1}{c}{$\tilde{A}^{\sin(\phi)}_{LU}$} & \multicolumn{1}{c|}{$A^{\sin(\phi)}_{LU}$}\\
 & & $\left[\text{GeV}\right]$ &  &  & $\left[\text{GeV}^2\right]$  &  & $\left[\text{GeV}\right]$  &  & \\ \hline
0.023 -- 0.071 & 0.20 -- 0.30 & 0.05 -- 0.23 & 0.047 & 0.681 & 1.639 & 0.259 & 0.167 & -0.0180 $\pm$ 0.0566 $\pm$ 0.0108 & -0.0249 $\pm$ 0.0838 $\pm$ 0.0164 \\
               &              & 0.23 -- 0.35 & 0.045 & 0.703 & 1.616 & 0.256 & 0.294 & 0.0034 $\pm$ 0.0451 $\pm$ 0.0050 & 0.0045 $\pm$ 0.0663 $\pm$ 0.0073 \\
               &              & 0.35 -- 0.51 & 0.044 & 0.712 & 1.610 & 0.254 & 0.431 & 0.0407 $\pm$ 0.0341 $\pm$ 0.0033 & 0.0579 $\pm$ 0.0499 $\pm$ 0.0049 \\
               &              & 0.51 -- 1.80 & 0.043 & 0.721 & 1.605 & 0.256 & 0.674 & -0.0072 $\pm$ 0.0264 $\pm$ 0.0008 & -0.0101 $\pm$ 0.0386 $\pm$ 0.0008 \\
               & 0.30 -- 0.37 & 0.05 -- 0.23 & 0.051 & 0.582 & 1.514 & 0.336 & 0.160 & -0.1102 $\pm$ 0.0491 $\pm$ 0.0066 & -0.1892 $\pm$ 0.0795 $\pm$ 0.0091 \\
\hline

 \end{tabular}
\caption{Three-dimensional lepton asymmetries (column 9) and virtual-photon asymmetries (column 10) for 
protons as a function of $x_B$, $z$, and $P_{h\perp}$ for data collected on a hydrogen target.
The bin intervals are given in the first three columns. The average values of the indicated kinematic variables are given in columns 4 to 8.
For the quoted uncertainties, the first is statistical, while the second is systematic.}
\label{tab:p_3d_hy}
\end{table*}

\begin{table*}[h]
\footnotesize
\centering
\begin{tabular}{|c c c |c c c c c|r r|}
\hline
\multicolumn{3}{|c|}{Kinematic bin} & \multicolumn{5}{|c|}{Average value} & \multicolumn{2}{|c|} {$\bar{p}$}\\\hline
 $x_{B}$ & $z$ & $P_{h\perp}$ & $\langle x_B \rangle$ & $\langle y \rangle$ & $\langle Q^2 \rangle$ & $\langle z \rangle$ & $\langle P_{h\perp} \rangle$ & \multicolumn{1}{c}{$\tilde{A}^{\sin(\phi)}_{LU}$} & \multicolumn{1}{c|}{$A^{\sin(\phi)}_{LU}$}\\
 & & $\left[\text{GeV}\right]$ &  &  & $\left[\text{GeV}^2\right]$  &  & $\left[\text{GeV}\right]$  &  & \\ \hline
0.023 -- 0.071 & 0.20 -- 0.30 & 0.05 -- 0.23 & 0.045 & 0.701 & 1.620 & 0.255 & 0.169 & -0.0556 $\pm$ 0.1188 $\pm$ 0.0129 & -0.0748 $\pm$ 0.1737 $\pm$ 0.0177 \\
               &              & 0.23 -- 0.35 & 0.044 & 0.724 & 1.647 & 0.252 & 0.294 & -0.0384 $\pm$ 0.0913 $\pm$ 0.0121 & -0.0606 $\pm$ 0.1333 $\pm$ 0.0186 \\
               &              & 0.35 -- 0.51 & 0.043 & 0.727 & 1.624 & 0.253 & 0.431 & -0.0639 $\pm$ 0.0686 $\pm$ 0.0143 & -0.0891 $\pm$ 0.1002 $\pm$ 0.0196 \\
               &              & 0.51 -- 1.80 & 0.043 & 0.733 & 1.628 & 0.255 & 0.668 & -0.0950 $\pm$ 0.0519 $\pm$ 0.0083 & -0.1383 $\pm$ 0.0759 $\pm$ 0.0135 \\
               & 0.30 -- 0.37 & 0.05 -- 0.23 & 0.049 & 0.611 & 1.548 & 0.334 & 0.162 & 0.2040 $\pm$ 0.1249 $\pm$ 0.0576 & 0.3344 $\pm$ 0.1939 $\pm$ 0.0865 \\
\hline
 \end{tabular}
\caption{Three-dimensional lepton asymmetries (column 9) and virtual-photon asymmetries (column 10) for 
anti-protons as a function of $x_B$, $z$, and $P_{h\perp}$ for data collected on a hydrogen target.
The bin intervals are given in the first three columns. The average values of the indicated kinematic variables are given in columns 4 to 8.
For the quoted uncertainties, the first is statistical, while the second is systematic.}
\label{tab:antip_3d_hy}
\end{table*}

\begin{table*}[h]
\footnotesize
\centering
\begin{tabular}{|c c c |c c c c c|r r|}
\hline
\multicolumn{3}{|c|}{Kinematic bin} & \multicolumn{5}{|c|}{Average value} & \multicolumn{2}{|c|} {$\pi^+$}\\\hline
 $x_{B}$ & $z$ & $P_{h\perp}$ & $\langle x_B \rangle$ & $\langle y \rangle$ & $\langle Q^2 \rangle$ & $\langle z \rangle$ & $\langle P_{h\perp} \rangle$ & \multicolumn{1}{c}{$\tilde{A}^{\sin(\phi)}_{LU}$} & \multicolumn{1}{c|}{$A^{\sin(\phi)}_{LU}$}\\
 & & $\left[\text{GeV}\right]$ &  &  & $\left[\text{GeV}^2\right]$  &  & $\left[\text{GeV}\right]$  &  & \\ \hline
0.023 -- 0.071 & 0.20 -- 0.30 & 0.05 -- 0.23 & 0.052 & 0.563 & 1.460 & 0.244 & 0.158 & 0.0095 $\pm$ 0.0103 $\pm$ 0.0006 & 0.0209 $\pm$ 0.0174 $\pm$ 0.0004 \\
               &              & 0.23 -- 0.35 & 0.048 & 0.617 & 1.503 & 0.244 & 0.292 & -0.0023 $\pm$ 0.0097 $\pm$ 0.0002 & -0.0024 $\pm$ 0.0155 $\pm$ 0.0018 \\
               &              & 0.35 -- 0.51 & 0.047 & 0.648 & 1.527 & 0.246 & 0.426 & 0.0210 $\pm$ 0.0087 $\pm$ 0.0016 & 0.0311 $\pm$ 0.0134 $\pm$ 0.0009 \\
               &              & 0.51 -- 1.80 & 0.045 & 0.689 & 1.569 & 0.253 & 0.632 & -0.0110 $\pm$ 0.0102 $\pm$ 0.0003 & -0.0158 $\pm$ 0.0153 $\pm$ 0.0010 \\
               & 0.30 -- 0.37 & 0.05 -- 0.23 & 0.053 & 0.535 & 1.432 & 0.332 & 0.157 & 0.0070 $\pm$ 0.0163 $\pm$ 0.0018 & 0.0140 $\pm$ 0.0282 $\pm$ 0.0048 \\
\hline
 \end{tabular}
\caption{Three-dimensional lepton asymmetries (column 9) and virtual-photon asymmetries (column 10) for 
positively charged pions as a function of $x_B$, $z$, and $P_{h\perp}$ for data collected on a deuterium target.
The bin intervals are given in the first three columns. The average values of the indicated kinematic variables are given in columns 4 to 8.
For the quoted uncertainties, the first is statistical, while the second is systematic.}
\label{tab:pi+_3d_de}
\end{table*}

\begin{table*}[h]
\footnotesize
\centering
\begin{tabular}{|c c c |c c c c c|r r|}
\hline
\multicolumn{3}{|c|}{Kinematic bin} & \multicolumn{5}{|c|}{Average value} & \multicolumn{2}{|c|} {$\pi^-$}\\\hline
 $x_{B}$ & $z$ & $P_{h\perp}$ & $\langle x_B \rangle$ & $\langle y \rangle$ & $\langle Q^2 \rangle$ & $\langle z \rangle$ & $\langle P_{h\perp} \rangle$ & \multicolumn{1}{c}{$\tilde{A}^{\sin(\phi)}_{LU}$} & \multicolumn{1}{c|}{$A^{\sin(\phi)}_{LU}$}\\
 & & $\left[\text{GeV}\right]$ &  &  & $\left[\text{GeV}^2\right]$  &  & $\left[\text{GeV}\right]$  &  & \\ \hline
0.023 -- 0.071 & 0.20 -- 0.30 & 0.05 -- 0.23 & 0.051 & 0.564 & 1.457 & 0.244 & 0.158 & -0.0072 $\pm$ 0.0111 $\pm$ 0.0038 & -0.0132 $\pm$ 0.0186 $\pm$ 0.0053 \\
               &              & 0.23 -- 0.35 & 0.048 & 0.619 & 1.499 & 0.243 & 0.292 & 0.0083 $\pm$ 0.0105 $\pm$ 0.0015 & 0.0111 $\pm$ 0.0167 $\pm$ 0.0027 \\
               &              & 0.35 -- 0.51 & 0.046 & 0.647 & 1.520 & 0.245 & 0.426 & -0.0004 $\pm$ 0.0094 $\pm$ 0.0001 & -0.0007 $\pm$ 0.0144 $\pm$ 0.0005 \\
               &              & 0.51 -- 1.80 & 0.045 & 0.691 & 1.566 & 0.252 & 0.635 & -0.0043 $\pm$ 0.0107 $\pm$ 0.0013 & -0.0076 $\pm$ 0.0160 $\pm$ 0.0016 \\
               & 0.30 -- 0.37 & 0.05 -- 0.23 & 0.053 & 0.533 & 1.422 & 0.332 & 0.157 & 0.0148 $\pm$ 0.0178 $\pm$ 0.0025 & 0.0331 $\pm$ 0.0309 $\pm$ 0.0049 \\
\hline
 \end{tabular}
\caption{Three-dimensional lepton asymmetries (column 9) and virtual-photon asymmetries (column 10) for 
negatively charged pions as a function of $x_B$, $z$, and $P_{h\perp}$ for data collected on a deuterium target.
The bin intervals are given in the first three columns. The average values of the indicated kinematic variables are given in columns 4 to 8.
For the quoted uncertainties, the first is statistical, while the second is systematic.}
\label{tab:pi-_3d_de}
\end{table*}

\begin{table*}[h]
\footnotesize
\centering
\begin{tabular}{|c c c |c c c c c|r r|}
\hline
\multicolumn{3}{|c|}{Kinematic bin} & \multicolumn{5}{|c|}{Average value} & \multicolumn{2}{|c|} {$K^+$}\\\hline
 $x_{B}$ & $z$ & $P_{h\perp}$ & $\langle x_B \rangle$ & $\langle y \rangle$ & $\langle Q^2 \rangle$ & $\langle z \rangle$ & $\langle P_{h\perp} \rangle$ & \multicolumn{1}{c}{$\tilde{A}^{\sin(\phi)}_{LU}$} & \multicolumn{1}{c|}{$A^{\sin(\phi)}_{LU}$}\\
 & & $\left[\text{GeV}\right]$ &  &  & $\left[\text{GeV}^2\right]$  &  & $\left[\text{GeV}\right]$  &  & \\ \hline
0.023 -- 0.071 & 0.20 -- 0.30 & 0.05 -- 0.23 & 0.052 & 0.565 & 1.469 & 0.247 & 0.156 & 0.0025 $\pm$ 0.0306 $\pm$ 0.0019 & 0.0265 $\pm$ 0.0505 $\pm$ 0.0048 \\
               &              & 0.23 -- 0.35 & 0.049 & 0.618 & 1.516 & 0.246 & 0.291 & 0.0400 $\pm$ 0.0302 $\pm$ 0.0060 & 0.0656 $\pm$ 0.0476 $\pm$ 0.0060 \\
               &              & 0.35 -- 0.51 & 0.047 & 0.642 & 1.525 & 0.247 & 0.426 & 0.0166 $\pm$ 0.0281 $\pm$ 0.0059 & 0.0245 $\pm$ 0.0430 $\pm$ 0.0049 \\
               &              & 0.51 -- 1.80 & 0.045 & 0.689 & 1.571 & 0.255 & 0.644 & 0.0123 $\pm$ 0.0304 $\pm$ 0.0088 & 0.0274 $\pm$ 0.0452 $\pm$ 0.0106 \\
               & 0.30 -- 0.37 & 0.05 -- 0.23 & 0.054 & 0.535 & 1.449 & 0.333 & 0.157 & 0.0238 $\pm$ 0.0430 $\pm$ 0.0023 & 0.0688 $\pm$ 0.0741 $\pm$ 0.0023 \\
\hline
 \end{tabular}
\caption{Three-dimensional lepton asymmetries (column 9) and virtual-photon asymmetries (column 10) for 
positively charged kaons as a function of $x_B$, $z$, and $P_{h\perp}$ for data collected on a deuterium target.
The bin intervals are given in the first three columns. The average values of the indicated kinematic variables are given in columns 4 to 8.
For the quoted uncertainties, the first is statistical, while the second is systematic.}
\label{tab:K+_3d_de}
\end{table*}

\begin{table*}[h]
\footnotesize
\centering
\begin{tabular}{|c c c |c c c c c|r r|}
\hline
\multicolumn{3}{|c|}{Kinematic bin} & \multicolumn{5}{|c|}{Average value} & \multicolumn{2}{|c|} {$K^-$}\\\hline
 $x_{B}$ & $z$ & $P_{h\perp}$ & $\langle x_B \rangle$ & $\langle y \rangle$ & $\langle Q^2 \rangle$ & $\langle z \rangle$ & $\langle P_{h\perp} \rangle$ & \multicolumn{1}{c}{$\tilde{A}^{\sin(\phi)}_{LU}$} & \multicolumn{1}{c|}{$A^{\sin(\phi)}_{LU}$}\\
 & & $\left[\text{GeV}\right]$ &  &  & $\left[\text{GeV}^2\right]$  &  & $\left[\text{GeV}\right]$  &  & \\ \hline
0.023 -- 0.071 & 0.20 -- 0.30 & 0.05 -- 0.23 & 0.051 & 0.572 & 1.478 & 0.245 & 0.156 & 0.0361 $\pm$ 0.0404 $\pm$ 0.0076 & 0.0564 $\pm$ 0.0665 $\pm$ 0.0099 \\
               &              & 0.23 -- 0.35 & 0.048 & 0.622 & 1.500 & 0.245 & 0.291 & -0.0242 $\pm$ 0.0385 $\pm$ 0.0142 & -0.0370 $\pm$ 0.0606 $\pm$ 0.0195 \\
               &              & 0.35 -- 0.51 & 0.047 & 0.651 & 1.535 & 0.247 & 0.427 & -0.0697 $\pm$ 0.0352 $\pm$ 0.0091 & -0.1079 $\pm$ 0.0537 $\pm$ 0.0127 \\
               &              & 0.51 -- 1.80 & 0.045 & 0.693 & 1.576 & 0.253 & 0.646 & 0.0312 $\pm$ 0.0370 $\pm$ 0.0077 & 0.0447 $\pm$ 0.0549 $\pm$ 0.0113 \\
               & 0.30 -- 0.37 & 0.05 -- 0.23 & 0.053 & 0.538 & 1.438 & 0.333 & 0.155 & -0.0331 $\pm$ 0.0578 $\pm$ 0.0029 & -0.0551 $\pm$ 0.0989 $\pm$ 0.0009 \\
\hline
 \end{tabular}
\caption{Three-dimensional lepton asymmetries (column 9) and virtual-photon asymmetries (column 10) for 
negatively charged kaons as a function of $x_B$, $z$, and $P_{h\perp}$ for data collected on a deuterium target.
The bin intervals are given in the first three columns. The average values of the indicated kinematic variables are given in columns 4 to 8.
For the quoted uncertainties, the first is statistical, while the second is systematic.}
\label{tab:K-_3d_de}
\end{table*}

\begin{table*}[h]
\footnotesize
\centering
\begin{tabular}{|c c c |c c c c c|r r|}
\hline
\multicolumn{3}{|c|}{Kinematic bin} & \multicolumn{5}{|c|}{Average value} & \multicolumn{2}{|c|} {$p$}\\\hline
 $x_{B}$ & $z$ & $P_{h\perp}$ & $\langle x_B \rangle$ & $\langle y \rangle$ & $\langle Q^2 \rangle$ & $\langle z \rangle$ & $\langle P_{h\perp} \rangle$ & \multicolumn{1}{c}{$\tilde{A}^{\sin(\phi)}_{LU}$} & \multicolumn{1}{c|}{$A^{\sin(\phi)}_{LU}$}\\
 & & $\left[\text{GeV}\right]$ &  &  & $\left[\text{GeV}^2\right]$  &  & $\left[\text{GeV}\right]$  &  & \\ \hline
 0.023 -- 0.071 & 0.20 -- 0.30 & 0.05 -- 0.23 & 0.046 & 0.685 & 1.629 & 0.259 & 0.167 & -0.0160 $\pm$ 0.0652 $\pm$ 0.0044 & -0.0139 $\pm$ 0.0961 $\pm$ 0.0073 \\
               &              & 0.23 -- 0.35 & 0.044 & 0.705 & 1.597 & 0.256 & 0.295 & 0.0343 $\pm$ 0.0537 $\pm$ 0.0133 & 0.0584 $\pm$ 0.0790 $\pm$ 0.0189 \\
               &              & 0.35 -- 0.51 & 0.044 & 0.710 & 1.591 & 0.253 & 0.432 & 0.0728 $\pm$ 0.0391 $\pm$ 0.0083 & 0.1036 $\pm$ 0.0573 $\pm$ 0.0117 \\
               &              & 0.51 -- 1.80 & 0.043 & 0.720 & 1.600 & 0.256 & 0.672 & -0.0103 $\pm$ 0.0317 $\pm$ 0.0115 & -0.0135 $\pm$ 0.0464 $\pm$ 0.0165 \\
               & 0.30 -- 0.37 & 0.05 -- 0.23 & 0.051 & 0.582 & 1.516 & 0.336 & 0.159 & 0.0425 $\pm$ 0.0582 $\pm$ 0.0060 & 0.0644 $\pm$ 0.0938 $\pm$ 0.0054 \\
\hline
 \end{tabular}
\caption{Three-dimensional lepton asymmetries (column 9) and virtual-photon asymmetries (column 10) for 
protons as a function of $x_B$, $z$, and $P_{h\perp}$ for data collected on a deuterium target.
The bin intervals are given in the first three columns. The average values of the indicated kinematic variables are given in columns 4 to 8.
For the quoted uncertainties, the first is statistical, while the second is systematic.}
\label{tab:p_3d_de}
\end{table*}

\begin{table*}[h]
\footnotesize
\centering
\begin{tabular}{|c c c |c c c c c|r r|}
\hline
\multicolumn{3}{|c|}{Kinematic bin} & \multicolumn{5}{|c|}{Average value} & \multicolumn{2}{|c|} {$\bar{p}$}\\\hline
 $x_{B}$ & $z$ & $P_{h\perp}$ & $\langle x_B \rangle$ & $\langle y \rangle$ & $\langle Q^2 \rangle$ & $\langle z \rangle$ & $\langle P_{h\perp} \rangle$ & \multicolumn{1}{c}{$\tilde{A}^{\sin(\phi)}_{LU}$} & \multicolumn{1}{c|}{$A^{\sin(\phi)}_{LU}$}\\
 & & $\left[\text{GeV}\right]$ &  &  & $\left[\text{GeV}^2\right]$  &  & $\left[\text{GeV}\right]$  &  & \\ \hline
0.023 -- 0.071 & 0.20 -- 0.30 & 0.05 -- 0.23 & 0.046 & 0.700 & 1.642 & 0.256 & 0.165 & 0.1579 $\pm$ 0.1283 $\pm$ 0.0290 & 0.2241 $\pm$ 0.1875 $\pm$ 0.0405 \\
               &              & 0.23 -- 0.35 & 0.044 & 0.724 & 1.640 & 0.253 & 0.295 & -0.0775 $\pm$ 0.1080 $\pm$ 0.0056 & -0.1168 $\pm$ 0.1575 $\pm$ 0.0061 \\
               &              & 0.35 -- 0.51 & 0.043 & 0.725 & 1.622 & 0.253 & 0.431 & -0.1602 $\pm$ 0.0774 $\pm$ 0.0139 & -0.2354 $\pm$ 0.1134 $\pm$ 0.0175 \\
               &              & 0.51 -- 1.80 & 0.043 & 0.734 & 1.614 & 0.255 & 0.665 & 0.0791 $\pm$ 0.0625 $\pm$ 0.0147 & 0.1140 $\pm$ 0.0911 $\pm$ 0.0217 \\
               & 0.30 -- 0.37 & 0.05 -- 0.23 & 0.049 & 0.615 & 1.551 & 0.334 & 0.161 & 0.1721 $\pm$ 0.1413 $\pm$ 0.0237 & 0.2572 $\pm$ 0.2204 $\pm$ 0.0358 \\
\hline
 \end{tabular}
\caption{Three-dimensional lepton asymmetries (column 9) and virtual-photon asymmetries (column 10) for 
anti-protons as a function of $x_B$, $z$, and $P_{h\perp}$ for data collected on a deuterium target.
The bin intervals are given in the first three columns. The average values of the indicated kinematic variables are given in columns 4 to 8.
For the quoted uncertainties, the first is statistical, while the second is systematic.}
\label{tab:antip_3d_de}
\end{table*}

\end{document}